\documentclass[fleqn,usenatbib]{mnras}

\usepackage[T1]{fontenc}
\usepackage{booktabs}

\DeclareRobustCommand{\VAN}[3]{#2}
\let\VANthebibliography\thebibliography
\def\thebibliography{\DeclareRobustCommand{\VAN}[3]{##3}\VANthebibliography}

\usepackage{graphicx}	
\usepackage{amsmath}	
\usepackage{amssymb}	
\usepackage{verbatim}
\usepackage[dvipsnames]{xcolor}
\usepackage{newtxtext,newtxmath}
\usepackage{threeparttable}
\usepackage{bm}

\newcommand{\CI}{C\,{\sc i}}
\newcommand{\CII}{C\,{\sc ii}}
\newcommand{\HI}{H\,{\sc i}}
\newcommand{\OI}{O\,{\sc i}}
\newcommand{\DI}{D\,{\sc i}}

\definecolor{green}{rgb}{0,0.4,0}
\newcommand{\SB}[1]{{\color{red} [SB:~ #1]}}

\newcommand{\ioffe}{Ioffe Institute, {Polyteknicheskaya 26}, 194021 Saint-Petersburg, Russia}

\title[\CI\ in diffuse ISM]{Neutral carbon in diffuse interstellar medium: abundance matching { with $\rm H_2$} for DLAs at high redshifts.}

\author[]{
S.~A.~Balashev,$^1$\thanks{E-mail: s.balashev@gmail.com},
D.~N.~Kosenko,$^1$ \\
$^1$ \ioffe \\
}

\date{Accepted 2023. Received 2023; in original form 2023}

\pubyear{2023}

\begin{document}
\label{firstpage}
\pagerange{\pageref{firstpage}--\pageref{lastpage}}
\maketitle

\begin{abstract}
We present the study of \CI/H$_2$ relative abundance in the diffuse cold neutral medium. Using the chemical and thermal balance model we calculated the dependence of \CI/H$_2$ on the main parameters of the medium: hydrogen number density, metallicity, strength of the UV field, and cosmic ray ionization rate (CRIR). We show that observed relative \CI\ and H$_2$ column densities in damped Lyman alpha systems (DLAs) at high redshifts can be reproduced within our model assuming the {typically} expected conditions in the diffuse cold neutral medium (CNM). Using the additional observed information the on metallicity, \HI\ column density, {and} excitation of \CI\ fine-structure levels, as well as temperature we estimated for a wide range metallicities in the CNM at high redshifts that CRIRs to be in {the} range from $\sim10^{-16}$ to $\rm few \times 10^{-15}\,\rm s^{-1}$, hydrogen number densities to be in range $\sim10 - 10^3$\,cm$^{-3}$, and UV field in range from $10^{-2}$ to $\rm few \times 10^2$ of Mathis field. We argue, that since the observed quantities used in this work are quite homogeneous {and} much less affected by the radiative transfer effects (in comparison {with e.g.} dissociation of HD and UV pumping of H$_2$ rotational levels) our estimates are quite robust against the assumption of the exact geometrical model of the cloud and local sources of the UV field.
\end{abstract}




\section{Introduction}

Carbon is one of the most important { elements} in the neutral phase of the local interstellar medium (ISM). Indeed, at metallicities close to solar, its singly ionized form, \CII, provides the principal source of the electrons (that determine the chemistry in the medium) and {is} one of the dominant cooling processes in diffuse regions, that {determines} the formation of the cold phase of neutral medium (CNM). In turn, in translucent and dense molecular regions carbon is mainly in \CI\ and/or CO, which also dominate the cooling of the medium. 
Therefore, it is essential to study carbon in CNM of ISM both observationally and theoretically. 

From observations, the most robust way to constrain the abundance and ionization state of carbon in CNM is an absorption line spectroscopy. However, while \CI\ is typically a sub-dominant form of carbon, regarding measured column densities, its abundance can be much more confidently constrained than \CII. Additionally, \CI\ in absorption always observed populating three fine-structure levels, which provide an important diagnostics of the physical conditions in the medium \cite[see e.g.][]{Jenkins2001,Silva2002,Srianand2005,Noterdaeme2007,Balashev2011,Jorgenson2010,Klimenko2020,Kosenko2023}. \CII\ abundance \citep[][]{Jenkins2011,RomanDuval2019} and excitation of its fine-structure level \citep{Balashev2022} can also be used for constraints on the physical state of the gas. However, the usage of \CII\ is much more limited than \CI\ due to the availability of less number of transitions and saturation of the absorption lines, although the incidence rate of \CII\ is much higher than \CI. In turn, while rotational excitation of CO can provide independent diagnostics of the physical conditions (Klimenko et al. in prep), observations of CO are much more limited than \CII\ and \CI, since the cross-section of the associated dense medium is quite small.

The \CI\ chemistry in diffuse ISM was comprehensively considered theoretically by e.g. \citealt[][]{Liszt2011,Wolfire2008}. However, these studies considered the observations obtained in our galaxy, i.e. examined only the local ISM. In turn, to date, there is a quite large number of observations of \CI\ obtained in high-redshift DLAs, i.e. remote galaxies. These observations are mostly associated with the low-metallicity gas ($Z \lesssim 0.3$), where one can expect the changes in the thermal \cite[see e.g.][]{Bialy2019,Balashev2022} and ionization balance \citep[see e.g.][]{Balashev2020,Balashev2021}{, which is important for the chemistry of \CI.} An important constituent of these changes are cosmic rays, which start to be a dominant heating source {of cold ISM} at $Z \lesssim 0.3$ \citep{Bialy2019} and govern {its} ionization state. Therefore \CI\ is {expected} to be sensitive to the cosmic rays ionization rate (CRIR) and hence can provide a way to measure it.

Previously, some indirect methods were also used to estimate the cosmic ray ionization rate based on the observations of the abundance of a several molecules that are sensitive to the degree of ionization of the medium. Most of the molecules used, such as $\rm H_3^+$, $\rm OH^+$, $\rm H_2O^+$, $\rm HD$ \citep[see e.g.][]{Hollenbach2012, Indriolo2012, Indriolo2015, Kosenko2021}, represent {diffuse and translucent molecular gas.}
These methods show that there is a large spread in the measured CRIR: from $10^{-17}$ to $\sim \rm few \times 10^{-15}$\,s$^{-1}$. Such a scatter can be caused both by the natural inhomogeneity of the CRIR in the interstellar medium, associated with the locality of sources and the effects of cosmic ray propagation, and by systematic effects in the methods of estimation used. Therefore independent estimations of CRIR are highly valuable. 


In this work, we present an independent formalism to calculate \CI\ abundance in the diffuse ISM (see Sect.~\ref{sec:model}) and its application on the sample of \CI\ measurements obtained in the high-redshift H$_2$-bearing DLAs. In particular, in comparison with previous studies \citep[e.g.][]{Wolfire2008,Liszt2011} we focus on the low metallicity gas, how it affects the \CI\ abundance, and the importance of the cosmic rays in \CI\ production (see Sect.~\ref{sec:vary}). We developed the method (presented in Sect.~\ref{sec:phys}) for estimating the main physical conditions in the medium, based on measured \CI\ and molecular hydrogen, H$_2$, column densities, the population of the \CI\ fine structure levels, and kinetic temperature (derived through the ortho-para ratio of H$_2$). A new important feature of this work that discriminates it from our previous studies \citep[e.g.][]{Klimenko2020,Kosenko2023} is the {consideration of} \CI\ abundance and constraints on CRIR that \CI\ provides coupled with explicit calculations of the thermal state of the gas. We applied this method to a sample of known \CI/H$_2$-bearing DLAs at high redshifts and compared it with {some} previous estimates (see Sect.~\ref{sec:phys_cond}).

\section{Model}
\label{sec:model}
In this Section we describe a formalism that was used to calculate relative \CI/$\rm H_2$ abundances, thermal state, and the excitation of \CI\ fine-structure levels. 
In the following, during the modelling, we paid attention to the following quantities: \HI\ and H$_2$ column densities, metallicity, the ortho-para ratio of H$_2$ (providing the measurements of the kinetic temperature), and the column densities on three \CI\ fine-structure levels. The latter provide the measurement of the total column density of \CI\ and hence the relative abundance of \CI\ to H$_2$. These quantities from the observations will be used to derive constraints on the main physical parameters of the model.

\subsection{Model global parameters}
\label{sec:parameters}
We considered a homogeneous (specified by the total hydrogen number density $n_{\rm H}^{\rm tot}$) medium with metallicity $Z$ { (denoted by default in decimal units)} that is exposed by the UV field, $\chi$ \citep[normalized to Mathis field,][]{Mathis1983} and cosmic rays with ionization rate $\zeta$ -- the primary ionization rate per hydrogen atom (measured in $\rm s^{-1}$). These are the main global parameters of interest. 

For each considered species $\rm X$ in the medium we denote the number and column density as $n_{\rm X}$ and $N_{\rm X}$, respectively\footnote{In the following, we assume that the number and column densities are measured in units cm$^{-3}$ and cm$^{-2}$, respectively.}. These quantities evidently are related through the integration along the line of sight, $N_{\rm X} = \int n_{\rm X} dl$. The column densities, are also observed quantities, and hence are the subject of the model calculation. For convenience, in the following text, we will use H$_2$ column density, $N_{\rm H_2}$, as a natural unit of measure of the {cloud thickness, instead of physical length}. The column density of the species $\rm X$ {can be calculated} as $N_{\rm X} = \int n_{\rm X} (N_{\rm H_2}) \frac{dl}{dN_{\rm H_2}} dN_{\rm H_2} = \int \frac{n_{\rm X}}{n_{\rm H_2}} dN_{\rm H_2}$.

\subsection{Metal and dust abundance}
\label{sec:metal_and_dust}
The thermal balance depends on the metal and dust abundances. The dust abundance was specified by the dust-to-gas ratio, $\rm DTG$. The observations suggest that it is scaled with metallicity as a power law or broken power law. However, the measured dispersion of DTG is quite large, therefore for simplicity, we assume that
\begin{equation}
\label{eq:dtg}
    [{\rm DTG}] \equiv \frac{{\rm DTG_{\,\,\,\,}}}{{\rm DTG}_{\odot}} = Z^{\beta},
\end{equation}
where ${\rm DTG}_{\odot}$ is the dust-to-gas ratio at solar abundance and $\beta$ is typically constrained between 1 and 3. While this dependence is simplistic, as was mentioned by \citealt{Balashev2022}, it matches with the observations from \cite{RemyRuyer2014}. Finally, when we used {the model calculations} we took into account the dispersion of DTG-metallicity dependence.

The abundance of the elements in the gas phase (which is also necessary to properly calculate the cooling rate) can be determined using the metallicity and DTG ratio as 
\begin{equation}
    [{\rm X/H}] \equiv \frac{{\rm X/H}_{\,\,\,\,}}{{\rm X/H}_{\odot}} = Z \left(1 - d_{\rm X} \cdot [{\rm DTG}]\right),
\end{equation}
where $\rm X/H_{\odot}$ is a solar photospheric abundance of {some species denoted by $X$} (we use values taken from the \citealt{Asplund2009}), and $d_{\rm X}$ is a depletion of species at solar metallicity. For the latter, we used the following values 0.53, 0.41, and 0.0 for carbon, oxygen, and argon, respectively. We did not apply depletion for other species, since they participate in neither thermal balance nor basic chemical reactions, considered in this work.

We also considered Polycyclic Aromatic Hydrocarbons (PAH) separately, while, we assumed that PAH abundance is linearly scaled with DTG ratio, i.e. the distribution of the grain size does not change with metallicity. In principle it doesn't have to be true, since the production of the large grains can be non-linearly dependent on the small grain abundance, as well as the large grains destructed/formed into/from  smaller grains, which may change their relative fractions. While other authors \cite{Wolfire2003, Wolfire2008, Bialy2019} used artificially added $\phi_{\rm PAH}$ coefficient to rescale the collisional coefficients of the PAH measured in the laboratory, we did not consider this factor, i.e. used $\phi_{\rm PAH}=1$.

\subsection{H$_2$/HI abundance}
\label{sec:HI_H2}

The abundances of H$_2$ and \HI\ determine the ionization state and chemistry of the cloud and are hence important for the thermal state and chemical abundances of the species. In principle, to calculate the relative H$_2$ and \HI\ abundances, and the \HI/$\rm H_2$ transition one needs to calculate full radiative transfer in H$_2$ Lyman and Werner bands. However, this is a very time-consuming task and here to be able to efficiently calculate the models, we used the analytical description of \HI/$\rm H_2$ transition proposed recently by \cite{Sternberg2014, Bialy2016}\footnote{We also applied it in our recent studies of $\rm HD$/$\rm H_2$ and $\rm OH$/$\rm H_2$ in \citealt{Balashev2020, Balashev2021}.}. Following this formalism, the number density of $\rm H_2$ as a function of $\rm H_2$ column density, $N_{\rm H_2}$ can be written as
\begin{equation}
    n_{\rm H_2} = \frac{n_{\rm H}^{\rm tot}}{\alpha S_{\rm H_2}(N_{\rm H_2}) e^{-\sigma_g (N_{\rm H} + 2N_{\rm H_2})}+2},
\end{equation}
where $N_{\rm H}$ is the atomic hydrogen column density, which can be derived as a function of the H$_2$ column density \citep[see][]{Bialy2016}, $S_{\rm H_2}$ is a self-shielding function, $\sigma_{\rm g}\approx 1.9 \times 10^{-21} [\rm DTG]$ (in cm$^2$) is the dust Lyman-Werner photon absorption cross-section per hydrogen atom, and $\alpha$ is the ratio of the rates of free space $\rm H_2$ photo-dissociation and $\rm H_2$ formation on the dust grains. In turn, atomic hydrogen density is simply defined as $n_{\rm H} = n^{\rm tot}_{\rm H} - 2 n_{\rm H_2}$.  

\subsection{Chemical abundances}
\label{sec:chem_abund}

For the purpose of this work, we considered a sample of species, that are abundant in diffuse clouds and affect the thermal and ionization balance (hence have an impact on the chemistry and excitation of \CI). The major species of the hydrogen network are $\rm H$, $\rm H_2$, $\rm H^+$,  $\rm H_2^+$, $\rm H_3^+$. For the oxygen molecular network we considered $\rm O$, $\rm O^+$, $\rm OH^+$, $\rm H_2O^+$, $\rm H_3O^+$, $\rm H_2O$, and $\rm OH$. We also calculated $\rm He$, $\rm He^+$, $\rm HeH^+$, $\rm Ar$, $\rm Ar^+$, and $\rm ArH^+$ that impact the oxygen molecular chemistry and electron (denoted by $e$) abundance. In the carbon network, we considered only $\rm C$ and $\rm C^{+}$ and did not append the network of C-bearing molecular species. This is reasonable since the model will be applied only for diffuse clouds, before the onset of CO, where the {abundances} of any other C-bearing {molecules} {are} very low and {affect} neither carbon chemistry nor thermal balance. We also explicitly considered PAH species at three ionization states: $\rm PAH^0$, $\rm PAH^{+}$, and $\rm PAH^{-}$, since they affect ion recombination, such as $\rm H^{+}$ and $\rm C^{+}$. 

For $\rm H$ and $\rm H_2$ we used analytical description of the abundances (see Sect.~\ref{sec:HI_H2}). We also used the fixed $\rm He$, $\rm O$, and $\rm Ar$ abundance (as described in Sect.~\ref{sec:metal_and_dust}, except $\rm He$ for which we used value 0.085 of hydrogen abundance, independent from metallicity), since they are dominant ionization form in cold diffuse medium. For other species, we calculated the chemical abundances by solving the matrix equations of the chemical network. We used the same reaction rates as provided by \citealt{Balashev2021} mostly compiled from UMIST database \citealt{McElroy2013}. For reaction rates of PAHs we used compilation by \citealt{Wolfire2008}. {We additionally considered the dependence of photoreactions on the dilution of the UV field, with coefficients taken from the recent compilation by \citep{Heays2017}}. To solve chemical equations we additionally assume that the total $\rm PAH$ abundance is $n_{\rm PAH} + n_{\rm PAH^{-}} + n_{\rm PAH^{+}} \equiv n_{\rm PAH}^{\rm tot} = 6\times10^{-7} [\rm DTG] n_{\rm H}^{\rm tot}$ and $n_{\rm C} + n_{\rm C^+} \equiv n_{\rm C}^{\rm tot} = [\rm C/H] n_{\rm H}^{\rm tot}$.

\subsection{Excitation of the fine-structure levels}
\label{sec:fine}

The excitation of fine-structure levels is important for the calculation of the cooling functions and the column densities of the fine-structure levels, that was used to compare with observed ones. We calculated the excitation of fine-structure levels of \CII, \CI\ and \OI\ using standard matrix calculations \citep[e.g.][]{Silva2002}, assuming excitation by collisions, {radiative excitation by CMB and UV pumping}, and de-excitation by spontaneous transitions. 
We considered collisions with the most abundant species in diffuse ISM:
 $\rm H$, $\rm H_2$ (separating ortho and para), $\rm He$ (if available){, and electrons}. The collisional coefficients for $\rm C^{+}$ were taken from \citep{Barinovs2005,Flower1977,Keenan1986}; for $\rm C$ from \citep{Abrahamson2007,Schroder1991,Staemmler1991,Johnson1987}; for $\rm O$ from  \citep{Abrahamson2007,Jaquet1992,Monteiro1987,Bell1998}. 
 We considered the excitation by Cosmic Microwave Background with the temperature set as $2.725 (1 + z)\rm\,K$ at the particular redshift of each studied absorber. This excitation can be important only for \CI\ (which has the smallest separation of the fine-structure levels) and redshift ($z\sim2-4$) while provides subdominant contribution to the {population of \CI\ levels, that is typically observed.}

\subsection{Thermal balance}
\label{sec:thermal}
The temperature of the medium was obtained using the thermal balance between the cooling and heating processes in ISM. We considered the cooling by Ly$\alpha$ \citep[see e.g.][]{Spitzer1978}, recombination \citep[we used coefficient from][]{Bialy2019}, and \CII, \CI, and \OI\ fine-structure emission. The latter was calculated using excitation of the fine-structure levels (see Sect.~\ref{sec:fine}) and assuming the optically thin regime. The main heating sources considered in this work are the photoelectric heating on the dust grains and cosmic rays, which depend on the physical parameters and electron abundance. We follow the standard formalism to obtain the heating rates, most recently described in \citealt{Bialy2019}. However, we additionally take into account the dependence of the photoelectric heating on the extinction of the UV field, using scaling of the rate by a dilution of the UV photons, calculated using the typical extinction curve \citep{Fitzpatrick2007} with $E_{B-V} = N_{\rm H}^{\rm tot} / 5.6\times10^{21} \rm [DTG]$.
We did not consider heating by X-ray background, since it is likely important only in the case of the presence of a strong X-ray emitting source. We also did not take into account the turbulence dissipation heating. While as was mentioned in \citep{Balashev2022} it can be important at low metallicities and even can be comparable with cosmic ray heating, the intermittent nature of the turbulence heating, may indicate that the heating can be very local \citep[see discussion in][]{Klessen2016}, which complicates the calculations. We also did not consider heating and cooling by $\rm H_2$, since it is only important either at low metallicity gas or at dense compact regions exposed {by} large UV field \cite[see e.g.][]{Bialy2019}, while we considered mostly diffuse ISM. 

\subsection{Solution }
\label{sec:solution}
To obtain the column densities of the species for each set of the physical parameters, defined in Sect.~\ref{sec:parameters} and \ref{sec:metal_and_dust}, we used the following routine. We divided the cloud of some length (specified by $N_{\rm H_2}$) by a number of layers, with 0.1 dex step starting from $\log N_{\rm H_2}=14$. For each layer, we first determine $n_{\rm H}$ and $n_{\rm H_2}$ using the analytical formalism of \HI/$\rm H_2$ transition (see Sect.~\ref{sec:HI_H2}). With these values and also fixed values of $\rm He$, $\rm O$, and $\rm Ar$ abundance (see Sect.~\ref{sec:chem_abund}) we calculate the cloud structure layer by layer (starting from the first) using iterative routine. At each iteration we solved a chemistry network to get abundances, $n_{\rm X}$ {and $N_{\rm X}$}, of other species in consideration (see Sect.~\ref{sec:chem_abund}) together with the thermal balance to get temperature (see Sect.~\ref{sec:thermal}) and excitation of fine-structure levels (see Sect.~\ref{sec:fine}). The iterations were stopped when next step does not change the derived values by more than 0.5 percent for each quantity, which set by a trade-off between accuracy and computational time. In most layers, several iterations were enough starting from the values, obtained at the previous layer. The typical calculation time of one model was about a few seconds per one core on the standard laptop, which we found quite enough {for} efficient sampling of the models presented in Sect.~\ref{sec:phys}.

\section{Data sample}
\label{sec:sample}
{To compare the model presented in the previous section with observations, we compiled observed \CI\ and $\rm H_2$ column densities from all known H$_2$-bearing DLAs detected at high redshift $(z\gtrsim 2)$ towards quasar sightlines. 
We note that there are a few additional detected systems at high-redshifts presented in \citep{Noterdaeme2018} that have only total \CI\ and H$_2$ column densities, i.e. there is no information on \CI\ fine-structure excitation, $T_{01}$ temperature, and metallicity. 
We also did not consider so-called proximate H$_2$-bearing DLAs \citep[e.g.][]{Noterdaeme2019,Noterdaeme2023}, which are typically defined as systems that have redshift close to the quasar redshift. i.e. $\Delta V \lesssim 3000\rm\,km/s$. These systems can be significantly impacted by the AGN emission \citep[e.g.][]{Balashev2020b} or even be produced in the AGN outflowing gas \citep[e.g.][]{Noterdaeme2021}, therefore they may be not representative for the diffuse medium, and require a separate study. 
The compiled sample is presented in Table~\ref{tab:obs}. In the case where several H$_2$/\CI-bearing components were detected along the line of sight, we provide explicitly the individual H$_2$/\CI\ components within DLAs, since the described modelling allows us to consider them independently. 
}

\begin{table*}
\caption{{Observational data on \CI/H$_2$-bearing absorption systems at high redshifts $(z\gtrsim 2)$ used in this work} 
}
\begin{tabular}{lccccccccc}
\toprule
       QSO & $z_{\rm abs}$ & $\log N_{\rm HI}$ & $\log N_{\rm H2}$ & $\log N_{\rm CIj0}$ & $\log N_{\rm CIj1}$ & $\log N_{\rm CIj2}$ & $\log Z$ & $T_{01}$$\,$[K] & References  \\
\midrule
J0000$+$0048 & 2.52546 & $20.80\pm 0.10$ & $20.43\pm 0.02$ & $16.10\pm 0.08$ & $15.54\pm 0.14$ & $14.67\pm0.11$ & $0.46\pm 0.45$ & $52\pm2$  & (1)     \\
B0027$-$1836 & 2.40183  & $21.75\pm0.10$ & $17.30\pm 0.07$ & $12.25^{+0.09}_{-0.15}$ & $<12.27$ & - & $-1.63\pm0.10$ & $134^{+41}_{-25}$ & (2, 3) \\
J0812$+$3208 & 2.626443 & $21.35\pm 0.10$ & $19.93\pm 0.04$ & $13.30\pm 0.23$ & $13.02\pm 0.03$ & $12.47\pm 0.05$ & $-0.81\pm 0.10$ & $48\pm 2$ & (4, 5)  \\
J0812$+$3208 & 2.626276 & '-' & $18.82\pm 0.37$ & $12.70\pm 0.02$ & $12.32\pm 0.04$  & $< 12.39$ & '-' & $50^{+44}_{-16}$ &  (4, 5)  \\
J0812$+$3208 & 2.625808 & '-' &  $15.98^{+0.29}_{-0.23}$$^{\dagger}$ & $12.13\pm0.05$ & $11.68\pm 0.16$ & $11.37\pm 0.27$ & '-' & $147^{+13}_{-11}$ & (4, 5)  \\
J0816$+$1446 & 3.287420 & $22.00\pm 0.20$ & $18.62^{+0.21}_{-0.17}$ & $13.43\pm 0.01$ & $13.24\pm 0.02$ & $12.47\pm 0.07$ & $-1.10\pm 0.10$ & $69^{+10}_{-8}$ & (6)    \\
J0843$+$0221$\ddagger$ & 2.7865 &  $21.82\pm 0.11$ & $21.21^{+0.02}_{-0.02}$ & $13.53^{+0.04}_{-0.02}$ & $13.61^{+0.02}_{-0.02}$ & $13.34^{+0.04}_{-0.04}$ &  $-1.52^{+0.08}_{-0.10}$ & $123^{+9}_{-8}$ & (7)      \\
J0906$+$0548 & 2.567181 & $20.13\pm 0.01$ & $18.88\pm 0.02$ & $13.50^{+0.08}_{-0.06}$ & $13.61^{+0.03}_{-0.03}$ & $12.89^{+0.07}_{-0.08}$ & $-0.16^{+0.05}_{-0.08}$ & $116^{+26}_{-4}$ & (8) \\
J0946$+$1216 & 2.606406 & $21.15\pm 0.02$ & $19.96^{+0.01}_{-0.02}$ & $14.10^{+0.05}_{-0.05}$ & $13.89^{+0.03}_{-0.04}$ & $13.28^{+0.05}_{-0.04}$ & $-0.46\pm 0.02$ & $131^{+10}_{-14}$ & (8)       \\
J0946$+$1216 & 2.607083 & '-' & $17.26^{+0.31}_{-0.08}$ & $13.22^{+0.05}_{-0.06}$ & $13.35^{+0.03}_{-0.03}$ & $12.83^{+0.10}_{-0.18}$ & '-' & $>90$ & (8)     \\
J1146$+$0743 & 2.839459 & $21.54\pm 0.01$ & $18.76^{+0.01}_{-0.01}$ & $13.28^{+0.05}_{-0.04}$ & $13.10^{+0.09}_{-0.10}$ & $12.93^{+0.12}_{-0.18}$ & $-0.57\pm 0.02$ & $91^{+3}_{-2}$ & (8)      \\
J1146$+$0743 & 2.841629 & '-' & $17.94^{+0.11}_{-0.13}$ & $13.17^{+0.10}_{-0.11}$ & $13.07^{+0.05}_{-0.08}$ & $12.85^{+0.12}_{-0.11}$ & '-' & $57^{+25}_{-16}$ & (8)      \\
Q1232$+$0815 & 2.33771 & $20.90\pm 0.09$ & $19.57\pm 0.11$ & $13.87\pm 0.05$ & $13.56\pm 0.04$ & $12.82\pm 0.07$ & $-1.32\pm 0.12$ & $67^{+20}_{-12}$ & (9, 10)       \\
J1237$+$0647 & 2.689550 & $20.00\pm 0.15$ & $19.20^{+0.13}_{-0.12}$ & $14.67\pm 0.04$ & $14.46\pm 0.03$ & $13.64\pm 0.02$ & $0.34\pm 0.12$ & $108^{+84}_{-33}$ & (11)       \\
J1237$+$0647 & 2.68801 & '-' & $16.28\pm0.10$ & $13.24\pm 0.04$ & $12.80\pm 0.03$ & - & '-' & $108^{+84}_{-33}$ & (11)       \\
J1237$+$0647 & 2.68868 & '-' & $17.62^{+0.08}_{-0.11}$ & $12.84\pm 0.03$ & $12.70\pm 0.05$ & - & '-' & $108^{+84}_{-33}$ & (11)       \\
J1311$+$2225 &3.091410 & $20.75\pm 0.10$ & $17.87^{+0.37}_{-0.33}$ & $13.17^{+0.11}_{-0.11}$ & $13.28^{+0.10}_{-0.11}$ & $12.00^{+0.5}_{-0.4}$ & $-0.34^{+0.13}_{-0.14}$ & $63^{+109}_{-27}$ & (12, 13)     \\
J1311$+$2225 & 3.091535 & '-' &  $19.52^{+0.02}_{-0.02}$ & $13.81^{+0.73}_{-0.27}$ & $13.45^{+0.11}_{-0.08}$ & $12.86^{+0.07}_{-0.11}$ & '-' & $94^{+11}_{-5}$ & (12, 13)   \\
J1311$+$2225 & 3.091735 & '-' &  $18.25^{+0.22}_{-0.39}$ & $13.48^{+0.05}_{-0.05}$ & $13.42^{+0.07}_{-0.08}$ & $13.14^{+0.04}_{-0.09}$ & '-' & $97^{+108}_{-46}$ & (12, 13)    \\
J1311$+$2225 & 3.091858 & '-' & $18.57^{+0.05}_{-0.09}$ & $13.32^{+0.31}_{-0.20}$ & $13.40^{+0.16}_{-0.09}$ & $12.81^{+0.10}_{-0.11}$ & '-' & $85^{+25}_{-13}$ & (12, 13)    \\
J1439$+$1117$\ddagger$ & 2.41837  & $20.10^{+0.10}_{-0.08}$ &
$19.38^{+0.10}_{-0.10}$ & $14.26^{+0.02}_{-0.02}$ & $14.02^{+0.02}_{-0.02}$ & $13.10^{+0.02}_{-0.02}$ & $0.16\pm 0.11$ & $107^{+33}_{-20}$ & (14, 15)  \\
J1513$+$0352 & 2.463622 & $21.82\pm 0.02$ & $21.31^{+0.01}_{-0.01}$ & $14.82^{+0.18}_{-0.18}$ & $14.60^{+0.06}_{-0.06}$ & $14.03^{+0.06}_{-0.06}$ & $-0.84\pm 0.23$ & $92^{+5}_{-6}$ & (16)     \\
J2100$-$0641 & 3.09145 & $21.05\pm 0.15$ & $18.76\pm 0.03$ & $12.57\pm 0.03$ & $12.31\pm 0.08$ & - & $-1.21\pm 0.15$ & $159^{+44}_{-29}$ &  (4, 17, 18)       \\
J2123$-$0050 & 2.059300 & $19.18\pm 0.15$ & $17.94\pm 0.01$ & $13.73\pm 0.02$ & $13.42\pm 0.03$ & $12.43\pm 0.02$ & $-0.19\pm 0.15$ & $139^{+9}_{-8}$ &  (12, 19, 20)       \\
J2123$-$0050 & 2.059550 & '-' & $15.16\pm 0.02$ & $12.91\pm 0.03$ & $12.74\pm 0.10$ & $12.16\pm 0.11$ & '-' & $650^{+250}_{-140}$ &  (12, 19, 20)  \\
J2140$-$0321 & 2.339900 & $22.40\pm 0.10$ & $20.13\pm 0.07$ & $13.03\pm 0.04$ & $13.20\pm 0.04$ & $13.02\pm 0.05$ & $-1.05\pm 0.13$ & $75^{+12}_{-9}$ & (21)      \\
B2318$-$1107 & 1.98888 & $20.68\pm0.03$ & $15.49\pm 0.03$ & $12.63^{+0.02}_{-0.02}$ & $12.30^{+0.04}_{-0.04}$ & - & $-0.85\pm 0.06$ & $188^{+34}_{-25}$ & (2, 3) \\
J2340$-$0053 & 2.054170 & $20.53\pm 0.15$ & $15.99\pm 0.04$ & $12.27^{+0.03}_{-0.04}$ & $11.97^{+0.08}_{-0.13}$ & $10.23^{+0.44}_{-0.22}$ & $-0.74\pm 0.16$ & $181^{+77}_{-25}$ & (4, 13) \\
J2340$-$0053 & 2.054291 & '-' & $15.24\pm 0.05$ & $12.17^{+0.04}_{-0.04}$ & $12.29^{+0.05}_{-0.05}$ & $12.06^{+0.07}_{-0.09}$ & '-' & $400^{+590}_{-170}$ & (4, 13) \\
J2340$-$0053 & 2.054528 & '-' & $17.11\pm 0.12$ & $13.51^{+0.04}_{-0.03}$ & $13.06^{+0.01}_{-0.01}$ & $12.00^{+0.40}_{-0.20}$ & '-' & $147^{+253}_{-54}$ & (4, 13) \\
J2340$-$0053 & 2.054610 & '-' & $18.27\pm 0.06$ & $13.20^{+0.05}_{-0.04}$ & $12.67^{+0.02}_{-0.03}$ & $12.20^{+0.50}_{-0.50}$ & '-' & $183^{+97}_{-50}$ & (4, 13) \\
J2340$-$0053 & 2.054723 & '-' & $18.14\pm 0.04$ & $13.29^{+0.01}_{-0.01}$ & $12.97^{+0.01}_{-0.01}$ & $12.19^{+0.06}_{-0.05}$ & '-' & $175^{+63}_{-31}$ & (4, 13) \\
J2340$-$0053 & 2.054995 & '-' & $16.43\pm 0.03$ & $12.34^{+0.05}_{-0.05}$ & $11.72^{+0.15}_{-0.22}$ & $10.72^{+0.31}_{-0.53}$ & '-' & $280^{+145}_{-59}$ & (4, 13) \\
J2340$-$0053 & 2.055140 & '-' & $17.44^{+0.04}_{-0.05}$ & $12.42^{+0.05}_{-0.02}$ & $12.18^{+0.06}_{-0.07}$ & $10.35^{+0.27}_{-0.27}$ & '-' & $175^{+59}_{-33}$ & (4, 13) \\
J2347$-$0051 & 2.587969 & $20.47\pm 0.01$ & $19.44\pm 0.01$ & $14.28^{+0.36}_{-0.37}$ & $13.70^{+0.11}_{-0.08}$ & $13.18^{+0.11}_{-0.16}$ & $-0.82^{+0.05}_{-0.03}$ & $76^{+2}_{-2}$ & (8)       \\
\bottomrule
\end{tabular}
\begin{tablenotes}
     \item References: (1) \cite{Noterdaeme2017}; (2) \cite{Noterdaeme2007}; (3) \cite{Noterdaeme2008a}; (4) \cite{Jorgenson2010}; (5) \cite{Balashev2010}; (6) \cite{Guimaraes2012}; (7) \cite{Balashev2017}; (8) \cite{Balashev2019}; (9) \cite{Ivanchik2010}; (10) \cite{Balashev2011}; (11) \cite{Noterdaeme2010}; (12) \cite{Noterdaeme2018}; (13) \cite{Kosenko2021}; (14) \cite{Srianand2008}; (15) \cite{Noterdaeme2008b};  (16) \cite{Ranjan2018}; (17) \cite{Balashev2015}; (18) \cite{Ivanchik2015}; (19) \cite{Klimenko2016}; (20) \cite{Milutinovic2010}; (21) \cite{Noterdaeme2015};  \\
     $\dagger$ independent analysis \\
     $\ddagger$ for these DLAs we used the total column densities, since $J=0,1$ H$_2$ lines are not resolved, and hence we have no independent measurements in several velocity components resolved in other species \citep[see][]{Balashev2017, Srianand2008}. \\
     $^{*}$ there is a typo in Table 2 of \citealt{Balashev2019}, with the correct value provided in Table A.8 of that paper. \\
    \end{tablenotes}
\label{tab:obs}
\end{table*}

\section{Results}
\label{sec:results}
In this section, we present the results of the modelling and {its application} to the observational data. 

\begin{figure*}
    \includegraphics[width=1\linewidth]{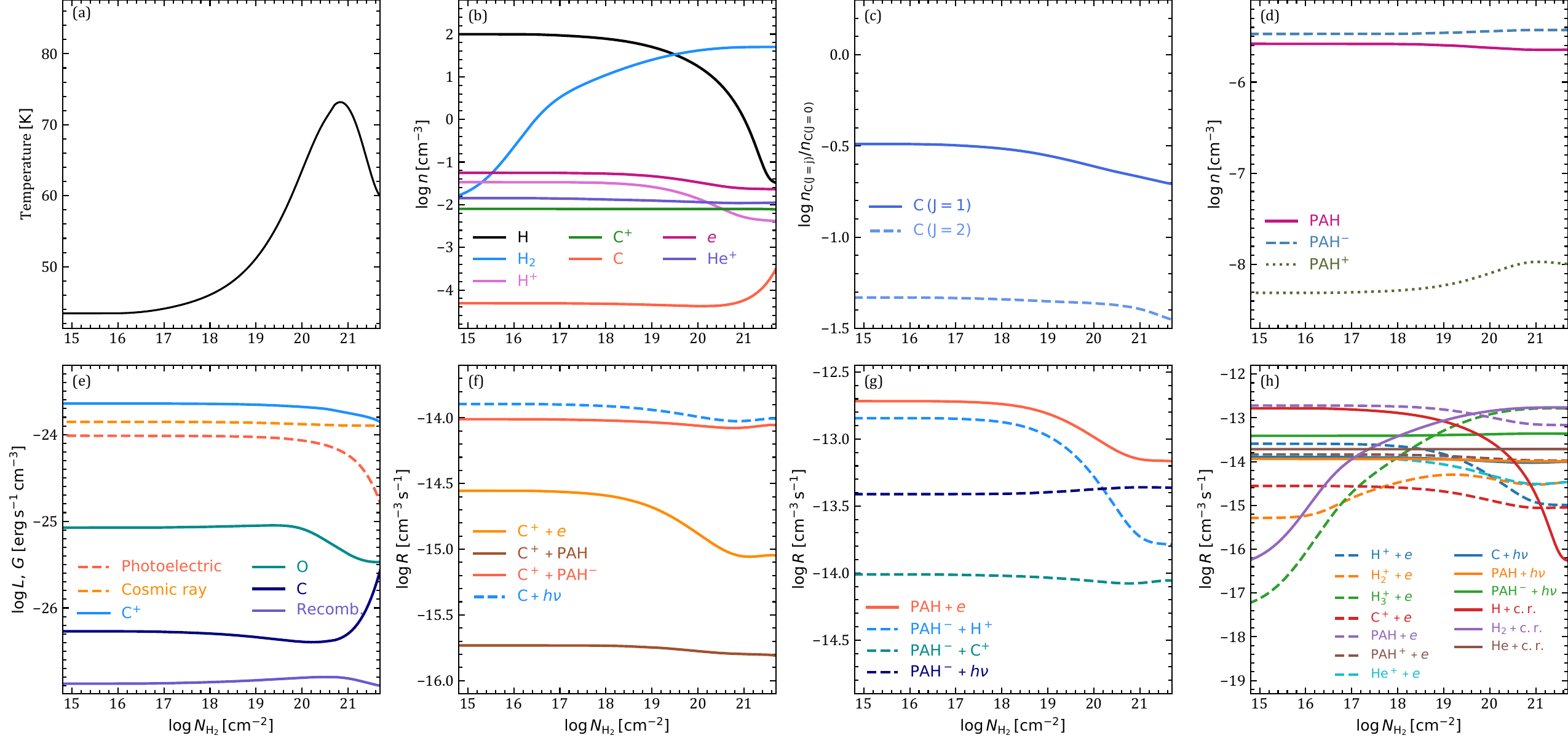}
    \caption{The results of the calculation of the cloud structure with $n_{\rm H}^{\rm tot} = 100\rm\,cm^{-3}$, $\chi = 1$, $\zeta=10^{-15}\rm\,s^{-1}$, and $Z=0.3$. The panels show profiles of: $\rm (a)$ temperature; $\rm (b)$ abundances of the main elements; $\rm (c)$ number densities on the $\rm C$ fine-structure levels; $\rm (d)$ PAH adundances; $\rm (e)$ the rate of the heating ({dashed} lines) and cooling ({solid} lines); $\rm (f,\,\,g,\,\,and\,\,h)$ the rate of reactions of the formation (solid lines) and destruction (dashed lines) for $\rm C$, $\rm PAH^{-}$, and $e$, respectively.}
    \label{fig:model}
\end{figure*}

An example of the calculated temperature, abundances, and excitation of \CI\ fine-structure levels, together with values of the chemical reaction and cooling/heating rates are shown in Fig.~\ref{fig:model} for the model with $n_{\rm H}^{\rm tot} = 100\rm\,cm^{-3}$, $\chi = 1$, $\zeta=10^{-15}\rm\,s^{-1}$, and $Z=0.3$. One can see that profiles of the species behave well as expected -- where the main feature is \HI/$\rm H_2$ transition (see panel $\rm (b)$ in Fig.~\ref{fig:model}), which affected the thermal balance (panel $\rm (e)$), temperature (panel $\rm (a)$), ionization (panels $\rm (b)$) and chemical structure (panel $\rm (g),\,(h)$) and at the less amount $\rm PAH$ abundances (panel $\rm (d)$). {It is } important to notice, that the temperature, \CI\ abundance, and relative population of \CI\ fine-structure levels are more or less constant and do not show strong variations throughout the cloud. Therefore these quantities are likely to have little bias regarding the geometrical model of the cloud, since they are little coupled with the radiative transfer. This will be important for the discussion of the results in the following subsections. {It is interesting to note that there is a slight increase of the kinetic temperature in the H$_2$ dominated region in comparison with atomic hydrogen dominated region, due to lower particle density in H$_2$ dominated region and lower collisional excitation rate of excited \CII\ fine-structure level by H$_2$, than by atomic hydrogen.}

\subsection{Variation of \CI/$\rm H_2$ abundance}
\label{sec:vary}
We investigate the dependence of \CI/$\rm H_2$ relative abundances on each physical parameter in the model. Since our calculations were based on one-sided radiation field models \citep[inherited from formalism by][]{Sternberg2014}, to compare model results with observational data, here and in the following text we calculated the column densities of each species at the half of specified $\rm H_2$ column density and then multiplied by two, i.e. $N_{\rm X}^{\rm obs} = 2 N_{\rm X}(N_{\rm H_2} / 2)$. This approximately emulates the plain parallel cloud of the thickness $N_{\rm H_2}$ exposed by UV field on both sides\footnote{It does not exactly correspond aforementioned case, since there is additional UV radiation coming from the opposite side, but since observed $N_{\rm H_2}$ column densities that we will consider are relatively high this will not introduce significant bias on the calculated quantities.}. 

\begin{figure*}
    \includegraphics[width=1\linewidth]{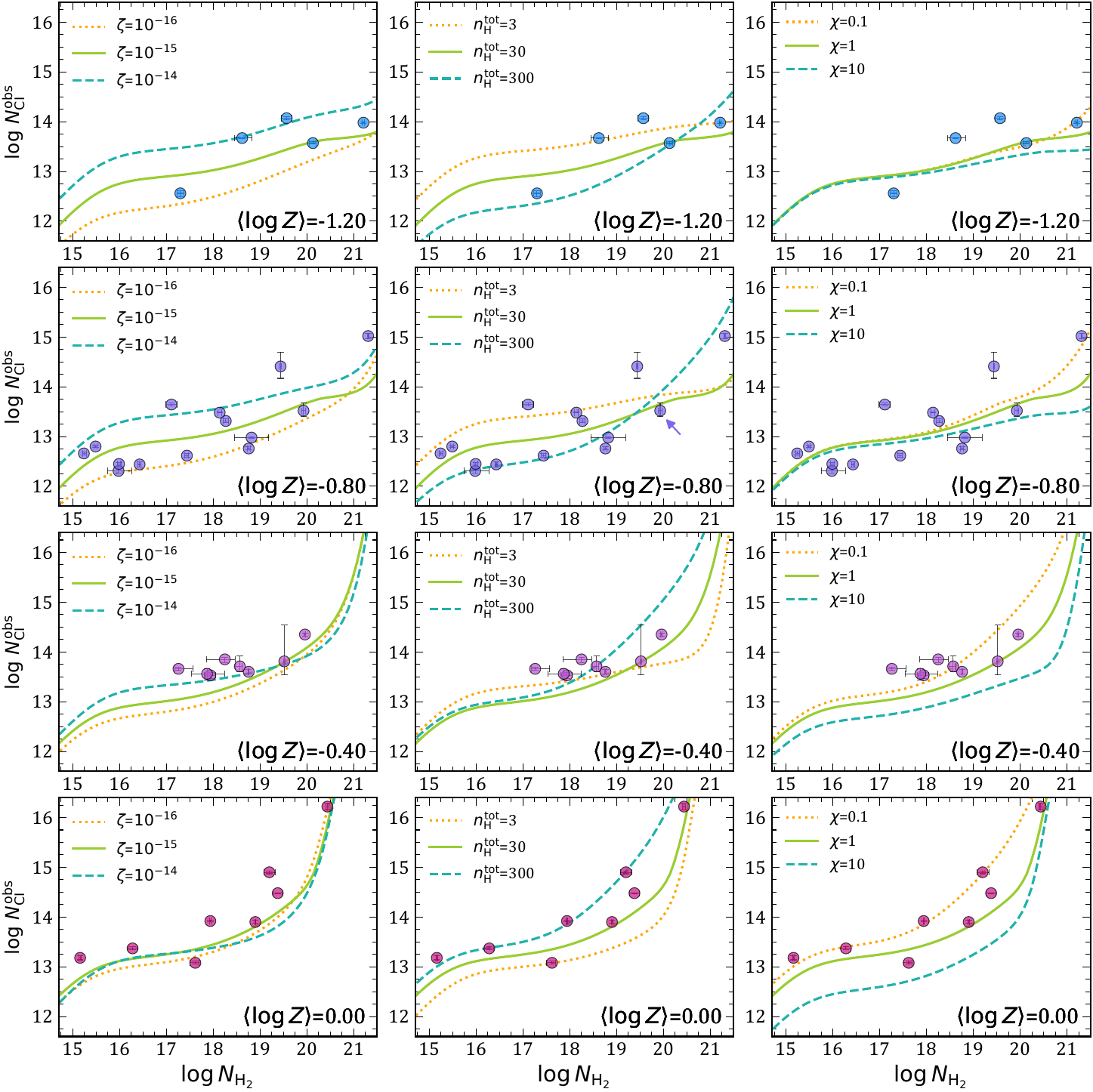}
    \caption{The dependence of the column densities of \CI\ on H$_2$. The colored points represent  observed abundances in high-redshift DLAs \citep[][and references therein]{Balashev2019,Klimenko2020,Kosenko2021} binned by the metallicity, where systems in each bin shown in rows, with mean metallicity within the bin provided in the right bottom corner of the panel. The calculated dependencies using the described formalism are presented by solid, dashed, and dotted lines, that correspond to the variation on one of the parameters from the base model with $n_{\rm H}^{\rm tot}=100$\,cm$^{-3}$, $\zeta=10^{-15}$\,s$^{-1}$ and $\chi=1$. {The arrow in the middle panel at the second row marked the position of J$\,$0812$+$3208, whose detailed analysis is presented in Figure.~\ref{fig:ind_res}}}
    \label{fig:CI_H2_var}
\end{figure*}

In Figure~\ref{fig:CI_H2_var} we plot the comparison of the observed and calculated dependence of \CI\ and $\rm H_2$ column densities. {We plot all the systems from the parent sample, presented in Sect.~\ref{sec:sample}, while we note that for several systems, where the column densities on high fine-structure were not measured, the total \CI\ column densities can be slightly higher, but not more than $\approx 30$ per cent. }
{It is evident to expect that \CI\ abundance may depend on the metallicity (at least carbon abundance almost linearly scaled with it), and since} the {observed metallicities at high redshifts systems} span about two orders from $\log Z \sim -1.5$ to $0.5$, we divided the sample in the four metallicity bins {$\approx0.4$ sizes (in dex)} and calculated theoretical $N_{\mbox{\CI}}^{\rm obs}(N_{\rm H_2})$ profiles for the mean values in each bin.

As we mentioned above, \CI\ abundance strongly depends on three global parameters of the model: the number density, $n_{\rm H}^{\rm tot}$, CRIR, $\zeta$, and UV field strength, $\chi$. To show how variations of these parameters affect \CI/$\rm H_2$ relative abundance we constructed the base model with $n_{\rm H}^{\rm tot}=100$\,cm$^{-3}$, $\zeta=10^{-15}\rm\,s^{-1}$ and $\chi=1$ and varied independently each of the parameters within two dexes, that correspond to the typical measured variations, see \cite{Klimenko2020,Kosenko2021}. The resulting profiles of \CI/H$_2$ abundances are shown in Figure~\ref{fig:CI_H2_var}. One can see that {except a few points at intermediate metallicities $\left<\log Z\right>\sim -0.8$ \textbf{and} $-0.4$ observed data can be reproduced} within chosen ranges of the physical parameters. 

One can see that at different metallicities \CI/H$_2$ abundance {is sensitive to different} parameters. At high metallicities, $Z\gtrsim 0.5$, it is mostly sensitive to the variation of the UV field and number density. This is simply because the UV field directly scales the \CI\ abundance by photoionization process. However, at low metallicities, $Z\lesssim 0.2$ (actually corresponding to the most DLAs at high redshifts), \CI/H$_2$ abundance also becomes quite sensitive to CRIR, since at these metallicities, \CII\ recombination rate depends on the electron abundance, which in turn depends on the hydrogen ionization fraction, which is sensitive to CRIR. Once metallicity approaches solar value, the electron densities start to be determined solely by the carbon abundance and hence \CI/H$_2$ becomes a little sensitive on CRIR (the bottom left panel in Figure~\ref{fig:CI_H2_var}). Interestingly,  at low metallicities, \CI/H$_2$ abundance becomes almost insensitive to the UV flux (the top central panel in Figure~\ref{fig:CI_H2_var}), at the typically observed 
values in the range between 0.1 and 10 of the Mathis field. 
{One can additionally note, that at intermediate H$_2$ column densities, i.e. $\log N_{\rm H_2} \lesssim 20$, \CI/H$_2$ abundance have opposite dependence on the number density at low and high metallicities, i.e. decreases and increases with the number density increase, respectively.}
    
\subsection{Constraints on the physical parameters}
\label{sec:phys}
{As we showed in previous section} abundance \CI/$\rm H_2$, \CI\ fine structure excitation and ortho-para ratio of H$_2$ depend on physical conditions ($n_{\rm H}^{\rm tot}$, $\chi$, $\zeta$) one can constrain the latter using observed quantities together with metallicity measurement. Importantly, as we {discussed} in Sect.~\ref{sec:results}, aforementioned {quantities (that we will use for comparison with observations) do not drastically vary along} the cloud, and hence these estimates are most probably less biased by unknown geometry of the absorbing region than most other species do, e.g. HD or O-bearing molecules. Since the dependency on the physical parameters likely has a relatively complex shape in the {parameter} space, and our model is quite a little time-consuming to constrain these parameters we followed the Bayesian approach to sample the posterior probabilities function of the $n_{\rm H}^{\rm tot}$, $\chi$, $\zeta$ using affine invariant Monte-Carlo Markov Chain sampler \citep{Goodman2010}. The likelihood function was constructed by comparison of model values with observations of the following quantities: (i) the kinetic temperature, measured by column densities of $J=0,1$ levels of H$_2$; (ii) the column densities of fine structure levels of \CI\, (this includes the relative excitation and total \CI\ column density) (iii) {the total \HI\ column density associated with DLA.} The latter { was} used as an upper limit only since we do not know how much of the observed \HI\ attributed to the \CI-bearing component, but we found it useful since it can shrink the posterior distribution functions. We considered H$_2$ column density as a model parameter since it determines the thickness of the cloud in the model. However, we take into account the uncertainty of the measurements of $N(\rm H_2)$ using a Gaussian prior, which is well within the Bayessian approach. Additionally, the metallicity was also considered as a model parameter with Gaussian prior corresponding to the measured interval estimate for each system. 
Since the \CI\ formation is quite sensitive to the dust-to-gas ratio, we considered the DTG ratio as an independent parameter during the modelling, while adding a prior to follow dependence with metallicity with {$\beta=-1.5$} (see Eq.~\ref{eq:dtg}) and additional 0.5 dex dispersion, to take into account the observed scatter in DTG-metallicity relation. For the other model parameters we used flat priors in log space, that conditionally emulate a wide distribution, {in the absence of other physically motivated priors}. In the following, to report the point and interval estimates on the model parameters we used maximum a posterior probability estimate and the highest posterior density 0.683 credible interval{, respectively}.

\begin{figure*}
    \includegraphics[width=1\linewidth]{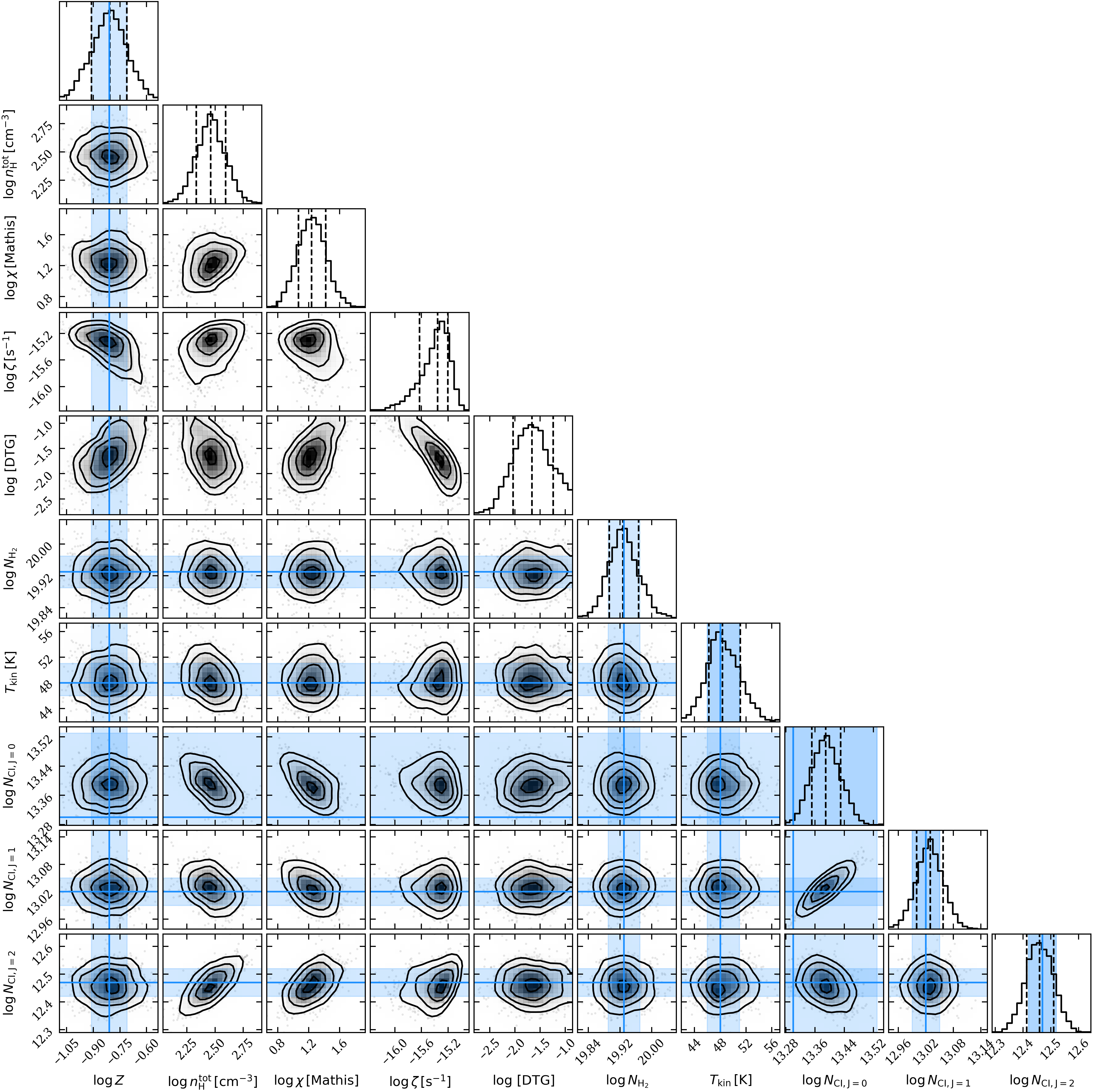}
    \caption{Posterior distribution functions of the constrained model parameters $n_{\rm tot}$, $\chi$, $\zeta$, and DTG {(along with the distributions of the measured quantities) for the H$_2$/\CI\ absorption system at z=2.626 towards Q\,0812+3208.} {The blue lines and stripes show the point and interval estimate from the observations.} The kinetic temperature, and column density of \CI\ fine-structure levels were used to calculate a likelihood function, while the total H$_2$ column density and metallicity were used as model parameters, with priors corresponding to the measured interval estimate.}
    \label{fig:ind_res}
\end{figure*}
An example of the result of the modelling of the H$_2$/\CI\ absorption system (at z=2.626 towards Q\,0812+3208) using described method is shown in Fig.~\ref{fig:ind_res} {using the corner plot python package \citep{corner}}. As we see in this particular system, we { were} able to reproduce {the observational quantities and obtain reasonable constraints} on the model parameters. In this particular system, we {found that} UV field, CRIR, and number density are relatively enhanced in comparison with typically { found} values for diffuse clouds. For the latter, it is consistent with the {observed} high excitation of \CI\ fine-structure levels.

\subsection{High redshift sample}
\label{sec:phys_cond}

We applied the procedure described in the previous Section to the \CI/H$_2$-bearing DLAs at high redshifts{, $z\gtrsim 2$,} detected so far. {The parent sample was described in Sect.~\ref{sec:sample} and shown in Table~\ref{tab:obs}. We selected a subsample of systems, where \CI\ column density on each fine-structure level was measured, to make the analysis homogeneous since excitation of \CI\ fine-structure levels provides important constraints on the model parameters.} 
The { described} modelling allows us to consider the individual H$_2$/\CI\ bearing components {of} DLAs, unless they have appropriate column density measurements of H$_2$, \CI\ fine-structure levels, and $T_{01}$ temperature. However, since we can not determine the fraction of the \HI\ associated with each component, we used an assumption that the metallicities of the components within DLA are the same and equal to the DLA average. 

The results of the {determinations} of the physical parameters are summarized in Table~\ref{tab:res}. We found that estimates on CRIR are located within the range from $10^{-16}$ to $\rm few \times 10^{-15}\,\rm s^{-1}$, while the UV estimates show a wider range of values. The number densities are found to be in the range of $n_{\rm H}^{\rm tot}=10-1000\,\rm cm^{-3}$ (note that this is the total density of hydrogen nuclei, so the actual number density can be $\approx$ two times lower than $n^{\rm tot}_{\rm H}$), which is consistent with the values, expected for the cold neutral medium.

\setlength{\tabcolsep}{4pt}
\begin{table}
\begin{tabular}{lcccc}
\toprule
       QSO &  $z_{\rm abs}$ &  $\log n_{\rm H}^{\rm tot} [\rm cm^{-3}]$ & $\log \chi [\rm Mathis]$ &          $\log \zeta [\rm s^{-1}]$ \\
\midrule
J0000+0048 &       2.525458 & $2.0^{+0.5}_{-0.4}$ &              $<-0.4$ &                  $<-14.8$ \\
J0812+3208 &       2.626443 & $2.5^{+0.1}_{-0.1}$ &  $1.2^{+0.2}_{-0.2}$ &     $-15.3^{+0.2}_{-0.2}$ \\
J0812+3208 &       2.625808 & $1.2^{+0.4}_{-0.4}$ &  $1.5^{+0.4}_{-0.4}$ &     $-16.2^{+0.6}_{-0.5}$ \\
J0816+1446 &       3.287420 & $2.1^{+0.3}_{-0.2}$ &              $<-1.6$ &     $-15.2^{+0.2}_{-0.2}$ \\
J0906+0548 &       2.569181 & $2.1^{+0.1}_{-0.1}$ &  $0.6^{+0.2}_{-0.3}$ &     $-15.0^{+1.1}_{-1.1}$ \\
J0946+1216 &       2.606406 & $2.1^{+0.1}_{-0.2}$ & $-0.1^{+0.7}_{-0.3}$ &     $-14.5^{+0.1}_{-0.2}$ \\
J0946+1216 &       2.607083 & $1.9^{+0.5}_{-0.4}$ &               $<0.7$ &     $-13.9^{+0.2}_{-0.2}$ \\
J1146+0743 &       2.839459 & $2.1^{+0.2}_{-0.1}$ &  $0.5^{+0.4}_{-0.8}$ &     $-14.4^{+0.2}_{-0.3}$ \\
J1146+0743 &       2.841629 &              $>2.7$ &  $1.4^{+0.5}_{-0.5}$ &                  $<-13.4$ \\
Q1232+0815 &       2.337710 & $2.3^{+0.1}_{-0.1}$ & $-2.3^{+0.4}_{-0.4}$ &     $-15.6^{+0.2}_{-0.2}$ \\
J1237+0647 &       2.689550 & $1.9^{+0.1}_{-0.1}$ & $-0.6^{+0.3}_{-0.5}$ &                  $<-15.3$ \\
J1311+2225 &       3.091410 &              $>1.7$ &  $0.8^{+0.9}_{-0.9}$ &                  $<-13.5$ \\
J1311+2225 &       3.091535 & $2.0^{+0.2}_{-0.3}$ &  $0.7^{+0.3}_{-0.6}$ &                  $<-14.3$ \\
J1311+2225 &       3.091735 &              $>1.8$ &  $0.8^{+0.4}_{-0.5}$ &                  $<-13.7$ \\
J1311+2225 &       3.091858 & $2.3^{+0.2}_{-0.4}$ &  $0.7^{+0.4}_{-0.7}$ &     $-14.6^{+0.7}_{-1.9}$ \\
J1439+1118 &       2.418370 & $1.8^{+0.3}_{-0.2}$ & $-0.1^{+0.2}_{-0.2}$ &     $-15.1^{+0.5}_{-0.8}$ \\
J1513+0352 &       2.463622 & $2.5^{+0.1}_{-0.2}$ & $-0.4^{+1.1}_{-0.5}$ &     $-14.8^{+0.3}_{-0.3}$ \\
J2123-0050 &       2.059300 & $1.5^{+0.1}_{-0.1}$ &              $<-0.0$ &                  $<-14.5$ \\
J2123-0050 &       2.059550 & $1.8^{+0.4}_{-0.3}$ & $-0.0^{+0.2}_{-0.5}$ &     $-14.3^{+0.6}_{-0.9}$ \\
J2140-0321 &       2.339900 & $2.9^{+0.3}_{-0.2}$ &  $1.6^{+0.3}_{-0.5}$ &     $-14.4^{+0.1}_{-0.2}$ \\
J2340-0053 &       2.054170 & $1.2^{+0.4}_{-0.4}$ &  $0.9^{+0.4}_{-0.4}$ &                  $<-16.2$ \\
J2340-0053 &       2.054291 &              $>1.9$ &               $<1.9$ &     $-14.1^{+0.3}_{-0.3}$ \\
J2340-0053 &       2.054528 & $1.5^{+0.7}_{-0.4}$ &              $<-0.0$ &     $-15.0^{+0.4}_{-1.4}$ \\
J2340-0053 &       2.054610 & $1.2^{+0.2}_{-0.2}$ &              $<-0.1$ &     $-15.6^{+0.1}_{-0.1}$ \\
J2340-0053 &       2.054723 & $1.6^{+0.2}_{-0.1}$ &               $<0.3$ &     $-14.8^{+0.1}_{-0.1}$ \\
J2340-0053 &       2.054995 & $1.0^{+0.5}_{-0.7}$ &               $<0.9$ &                  $<-16.7$ \\
J2340-0053 &       2.055140 & $1.5^{+0.4}_{-0.3}$ &  $0.8^{+0.4}_{-0.4}$ &                  $<-16.0$ \\
J2347-0051 &       2.587969 & $2.4^{+0.4}_{-0.4}$ & $-1.2^{+0.7}_{-0.6}$ &     $-14.8^{+0.4}_{-0.3}$ \\
\bottomrule
\end{tabular}
\caption{The constraints on the physical conditions in high redshift \CI/H$_2$-bearing DLAs obtained from the joint modelling of \CI/H$_2$/\HI\ column densities, excitation of \CI\ fine-structure, and lower H$_2$ rotational levels.}
\label{tab:res}
\end{table}

\begin{figure}
    \includegraphics[width=1.0\linewidth]{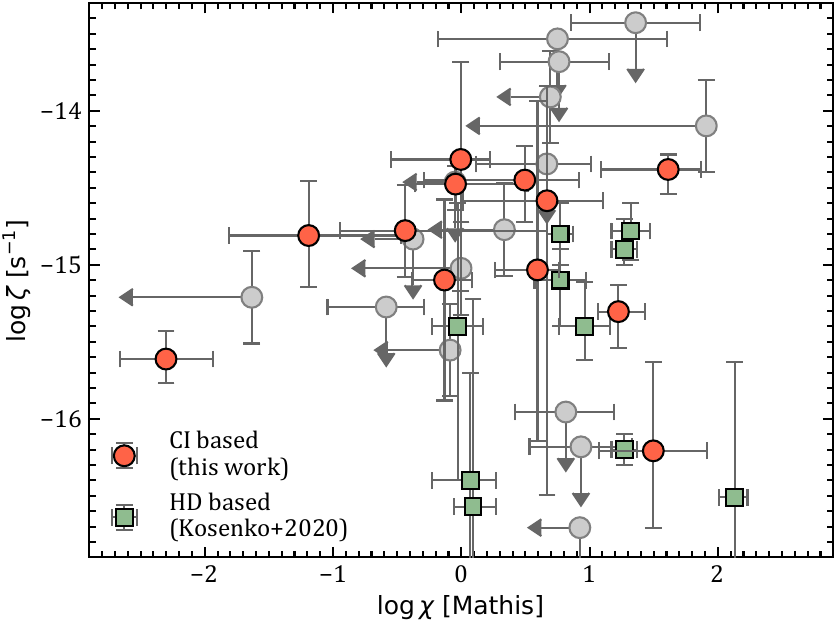}
    \caption{The { estimations} of cosmic ray ionization rate, $\zeta$, and UV field, $\chi$ in diffuse CNM at high redshift H$_2$-bearing DLAs. The red circles and green squares represent the { results} obtained based on \CI/H$_2$ (this work) and HD/H$_2$ \citep[from][]{Kosenko2021} abundances, respectively. The gray symbols depicted the upper limits.}
    \label{fig:crir_uv}
\end{figure}

In Fig.~\ref{fig:crir_uv} we show the relative dependence between { obtained} values of CRIR and UV fluxes. It seems that there is no evident correlation between these parameters unless one considers some fraction of the data. We note, that since these parameters compete on their contribution to the thermal balance (through cosmic ray heating or photoelectric heating) in some absorption systems we found that they are degenerated. However, usage of the \CI\ abundance can relax this degeneration, since it { provides} independent { constraints} on CRIR. Additionally, we did not confirm the quadratic dependence of the CRIR on UV flux, which was previously reported by \citealt{Kosenko2021} for the diffuse ISM, based on the modelling of HD/H$_2$ relative abundance coupled with \CI\ fine-structure excitation. In comparison with the aforementioned { values}, we obtained relatively higher values CRIR on average, while the significant fraction of the points around $\log \zeta \sim -15$ and $\log \chi \sim 1$ are consistent with each other. The higher values of the CRIR that we found from \CI\ modelling likely arose from the matching of \CI/H$_2$ abundance, since the \CI\ abundance depends on the recombination rate, which is tightly related to the electron abundance (directly at low metallicities, or indirectly through the dependence on the PAH$^-$ abundance). However, we note that the { constraints} based on \CI\ are more robust, regarding the radiative transfer issues than HD/H$_2$-based measurements. Indeed, \DI/HD transition which is important to the total HD abundance is typically sharp \citep[see e.g.][]{Balashev2020}, and hence can be sensitive to the exact geometrical model of the cloud, since the irregular shape, non-uniformity, and patchy structure may allow deeper penetration of the dissociating UV radiation in the medium and making HD abundance less than in case of uniform model. In that sense, one can consider that HD/H$_2$ likely provides a lower limit on the CRIR, since higher ionization fraction\footnote{As discussed in \citealt{Balashev2020}, HD is mainly formed through the reaction of $\rm H_2$ and $\rm D^{+}$, where abundance of the latter is tightly linked with $\rm H^{+}$ abundance, and hence the HD formation rate scales with the hydrogen ionization fraction in the medium and CRIR.} (and hence higher CRIR) is necessary if \DI/HD transition is shifted deeper into the cloud. Conversely, if we assume higher CRIR (and higher ionization fraction) \DI/HD transition occurs at a lower depth in the cloud. In turn, as we note above, all, \CI/H$_2$ abundance, \CI\ fine-structure excitation and the temperatures used in our study are quite homogeneous within the medium and do not reveal large variations that can significantly bias results.

\begin{figure}
    \includegraphics[width=1.0\linewidth]{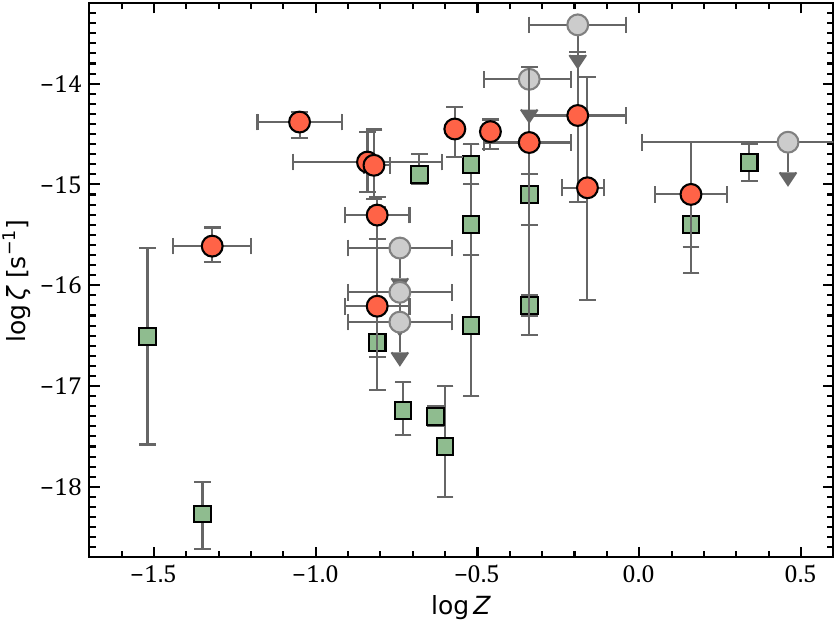}
    \caption{The { estimates} of cosmic ray ionization rate, $\zeta$, as a function of average metallicity in high redshift H$_2$-bearing DLAs. The color code is the same as in Fig.~\ref{fig:crir_uv}.}
    \label{fig:crir_z}
\end{figure}

As we discussed above, the ionization and the thermal state depend on the dust abundance, since the high metallicities, $\bm{\log}Z\gtrsim -0.5$, the photoelectric heating is the dominant heating source and the electron and H$^+$ recombination are tightly coupled with the PAH abundances. In turn, at low metallicities, $\bm{\log}Z\lesssim -0.5$, dust (as it scales with metallicity) is less important for both the ionization and thermal state and they are mostly governed by cosmic rays. Additionally, at low metallicities, the electron production rate becomes insensitive to metal abundance, since it is dominated by hydrogen ionization, rather than carbon photoionization as at high metallicity. Since our sample probes a large range of metallicities, we have an opportunity to explore this metallicity dependence. In Fig.~\ref{fig:crir_z} we plot the dependence of the derived CRIR on the metallicity. One can see that our { results} do not show any significant correlation with metallicity, which may indicate that the results are robust against artificially induced effects regarding the onset of the cosmic ray dominance at low metallicities. However, in principle, this cannot be separated from the possible dependence of the physical conditions on the metallicities, which is expected since different metallicities attributed to different galaxies and their evolutionary states. Indeed it is known, that the metallicity of the DLAs correlates with the mass of the galaxy and its halo. The previous { constraints} based on HD/H$_2$ abundance also do not reveal a strong correlation, although as we discussed above, have systematically lower values of the derived CRIR.

\begin{figure}
    \includegraphics[width=1.0\linewidth]{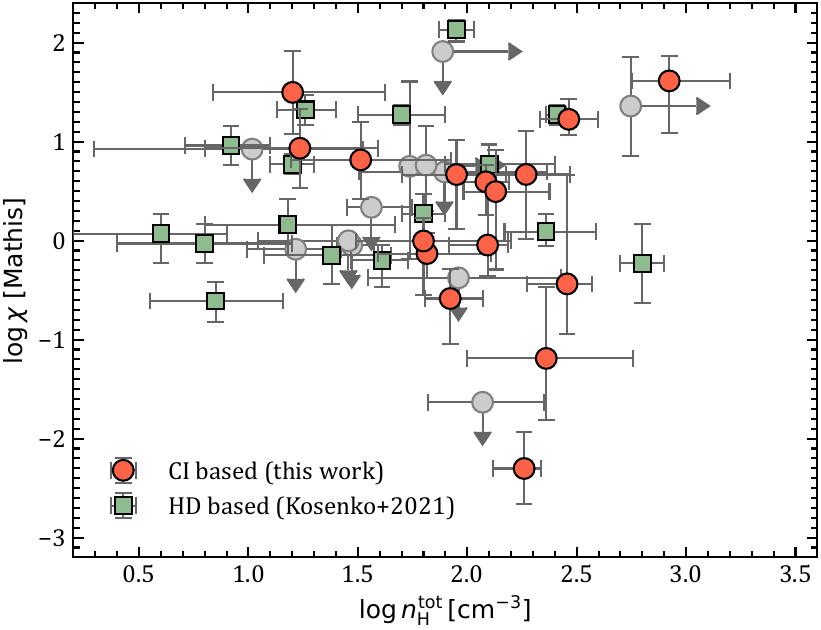}
    \caption{The { estimates} of UV field, $\chi$, and hydrogen number density, $n^{\rm tot}_{\rm H}$, in diffuse CNM at high redshift H$_2$-bearing DLAs. The color code is the same as in Fig.~\ref{fig:crir_uv}. 
    }
    \label{fig:uv_ntot}
\end{figure}

In Fig.~\ref{fig:uv_ntot} we plot the dependence of the derived UV flux on the hydrogen number density. One can see {again, that there is no strong} correlation between these quantities. 
However, one can notice that while UV field estimates are found to be consistent, we derive systematically higher values of the hydrogen number density in our study, than were previously obtained using only excitation of \CI\ fine-structure levels and kinetic temperature (note that values found by \citealt{Kosenko2021} generally agree with what was obtained by \citealt{Klimenko2020}). The main discrepancy can arise from the fact that we also used \CI\ abundance measurements and \CI\ formation rate scales with number density, and hence the higher number density is favored to reproduce the observed relatively high \CI/H$_2$ ratios. Additionally, we note, the results obtained by \citealt{Kosenko2021,Klimenko2020} based on the modelling by {\textsc MEUDON PDR} code, during which they used a fixed CRIR to minimize the number of the model variables. As we discussed above, the CRIR may significantly impact the thermal and chemical balance at low metallicities and therefore can elevate the characteristic number density of the CNM, since it affects the position of CNM branch of the neutral phase diagram \citep[see e.g.][]{Bialy2019,Balashev2022}. 
Additionally, as was noted by Slava Klimenko (private communication, Klimenko et al. in prep), the {previous} version of \textsc{MEUDON PDR} code {that was} used by \citealt{Klimenko2020,Kosenko2021} {had} incorrect treatment of CMB field, which intensity was a factor of 2 higher than it should be. Higher values of the CMB field, will increase the contribution of CMB to the excitation of \CI\ levels at high redshifts, where the CMB temperature is enhanced by (1+z) factor, hence lowering the number density estimates. {The detailed comparison of \CI/H$_2$ and HD/H$_2$ modelling certainly deserves an additional study.} {However, taking into account the aforementioned issues,} we suppose that the results presented in that study based on the \CI/H$_2$ abundance are more robust, than the previous ones \citep{Klimenko2020, Kosenko2021} based on {analysis with} \textsc{MEUDON PDR} code.

\section{Conclusions}

We present the modelling of the relative abundance of \CI/H$_2$ in the diffuse cold neutral medium using the solution of the \HI/H$_2$ transition, chemical network (with an explicit treatment of the PAH), and excitation of fine-structure levels coupled with the thermal balance. This allows us to accurately describe the dependence of \CI/H$_2$ on the global physical parameters, metallicity, number density, UV field, and cosmic ray ionization rate. We show that typically estimated values of these parameters can reproduce the observation from \CI/H$_2$-bearing DLA systems at high redshifts. These systems mostly represent the low-metallicity gas, for which the cosmic rays start to dominate the heating and ionization state of the medium, and hence severely affects the chemistry of the gas. We used \CI/H$_2$ abundances, the population of \CI\ fine structure, and H$_2$ rotational levels measured in DLAs to get constraints on global physical parameters. We found for the sample of all known \CI/H$_2$-bearing DLAs at high redshifts that CRIR is in the range $\sim10^{-16}$ to $\rm few \times 10^{-15}\,\rm s^{-1}$, hydrogen number density is in the range $\sim10 - 10^3$\,cm$^{-3}$, and UV field is in the range $10^{-2}$ to $\rm few \times 10^2$ of Mathis field. We did not find any strong correlation between {obtained} parameters.
We also compare {measured} values with previous {constraints} obtained using only excitation of \CI\ and H$_2$ levels, and HD/H$_2$ relative abundance. We argue that {the method proposed in this paper provides} more robust estimates {on the physical conditions, since it takes} into account the importance of CRIR and {is} based on the observational quantities, which are quite homogeneous within the medium, i.e. less affected by the radiative transfer effect, as e.g. HD/H$_2$ do. 

\section*{Data Availability}

The data used in this paper in available appropriate references provided in the text. The results of the calculations can be shared on request to the corresponding author.


\section{Acknowledgements} 
This work was supported by RSF grant 22-22-00164. We thank the anonymous referees for their valuable comments and useful suggestions.



\bibliographystyle{mnras}
\bibliography{references} 

\begin{thebibliography}{}
\makeatletter
\relax
\def\mn@urlcharsother{\let\do\@makeother \do\$\do\&\do\#\do\^\do\_\do\%\do\~}
\def\mn@doi{\begingroup\mn@urlcharsother \@ifnextchar [ {\mn@doi@}
  {\mn@doi@[]}}
\def\mn@doi@[#1]#2{\def\@tempa{#1}\ifx\@tempa\@empty \href
  {http://dx.doi.org/#2} {doi:#2}\else \href {http://dx.doi.org/#2} {#1}\fi
  \endgroup}
\def\mn@eprint#1#2{\mn@eprint@#1:#2::\@nil}
\def\mn@eprint@arXiv#1{\href {http://arxiv.org/abs/#1} {{\tt arXiv:#1}}}
\def\mn@eprint@dblp#1{\href {http://dblp.uni-trier.de/rec/bibtex/#1.xml}
  {dblp:#1}}
\def\mn@eprint@#1:#2:#3:#4\@nil{\def\@tempa {#1}\def\@tempb {#2}\def\@tempc
  {#3}\ifx \@tempc \@empty \let \@tempc \@tempb \let \@tempb \@tempa \fi \ifx
  \@tempb \@empty \def\@tempb {arXiv}\fi \@ifundefined
  {mn@eprint@\@tempb}{\@tempb:\@tempc}{\expandafter \expandafter \csname
  mn@eprint@\@tempb\endcsname \expandafter{\@tempc}}}

\bibitem[\protect\citeauthoryear{{Abrahamsson}, {Krems}  \&
  {Dalgarno}}{{Abrahamsson} et~al.}{2007}]{Abrahamson2007}
{Abrahamsson} E.,  {Krems} R.~V.,   {Dalgarno} A.,  2007, \mn@doi [\apj]
  {10.1086/509631}, \href
  {https://ui.adsabs.harvard.edu/abs/2007ApJ...654.1171A} {654, 1171}

\bibitem[\protect\citeauthoryear{{Asplund}, {Grevesse}, {Sauval}  \&
  {Scott}}{{Asplund} et~al.}{2009}]{Asplund2009}
{Asplund} M.,  {Grevesse} N.,  {Sauval} A.~J.,   {Scott} P.,  2009, \mn@doi
  [\araa] {10.1146/annurev.astro.46.060407.145222}, \href
  {https://ui.adsabs.harvard.edu/abs/2009ARA&A..47..481A} {47, 481}

\bibitem[\protect\citeauthoryear{{Balashev} \& {Kosenko}}{{Balashev} \&
  {Kosenko}}{2020}]{Balashev2020}
{Balashev} S.~A.,  {Kosenko} D.~N.,  2020, \mn@doi [\mnras]
  {10.1093/mnrasl/slz180}, \href
  {https://ui.adsabs.harvard.edu/abs/2020MNRAS.492L..45B} {492, L45}

\bibitem[\protect\citeauthoryear{{Balashev}, {Ivanchik}  \&
  {Varshalovich}}{{Balashev} et~al.}{2010}]{Balashev2010}
{Balashev} S.~A.,  {Ivanchik} A.~V.,   {Varshalovich} D.~A.,  2010, \mn@doi
  [Astronomy Letters] {10.1134/S1063773710110010}, \href
  {https://ui.adsabs.harvard.edu/abs/2010AstL...36..761B} {36, 761}

\bibitem[\protect\citeauthoryear{{Balashev}, {Petitjean}, {Ivanchik}, {Ledoux},
  {Srianand}, {Noterdaeme}  \& {Varshalovich}}{{Balashev}
  et~al.}{2011}]{Balashev2011}
{Balashev} S.~A.,  {Petitjean} P.,  {Ivanchik} A.~V.,  {Ledoux} C.,  {Srianand}
  R.,  {Noterdaeme} P.,   {Varshalovich} D.~A.,  2011, \mn@doi [\mnras]
  {10.1111/j.1365-2966.2011.19489.x}, \href
  {https://ui.adsabs.harvard.edu/abs/2011MNRAS.418..357B} {418, 357}

\bibitem[\protect\citeauthoryear{{Balashev}, {Noterdaeme}, {Klimenko},
  {Petitjean}, {Srianand}, {Ledoux}, {Ivanchik}  \& {Varshalovich}}{{Balashev}
  et~al.}{2015}]{Balashev2015}
{Balashev} S.~A.,  {Noterdaeme} P.,  {Klimenko} V.~V.,  {Petitjean} P.,
  {Srianand} R.,  {Ledoux} C.,  {Ivanchik} A.~V.,   {Varshalovich} D.~A.,
  2015, \mn@doi [\aap] {10.1051/0004-6361/201425553}, \href
  {https://ui.adsabs.harvard.edu/abs/2015A&A...575L...8B} {575, L8}

\bibitem[\protect\citeauthoryear{{Balashev} et~al.,}{{Balashev}
  et~al.}{2017}]{Balashev2017}
{Balashev} S.~A.,  et~al., 2017, \mn@doi [\mnras] {10.1093/mnras/stx1339},
  \href {https://ui.adsabs.harvard.edu/abs/2017MNRAS.470.2890B} {470, 2890}

\bibitem[\protect\citeauthoryear{{Balashev} et~al.,}{{Balashev}
  et~al.}{2019}]{Balashev2019}
{Balashev} S.~A.,  et~al., 2019, \mn@doi [\mnras] {10.1093/mnras/stz2707},
  \href {https://ui.adsabs.harvard.edu/abs/2019MNRAS.490.2668B} {490, 2668}

\bibitem[\protect\citeauthoryear{{Balashev}, {Ledoux}, {Noterdaeme},
  {Srianand}, {Petitjean}  \& {Gupta}}{{Balashev} et~al.}{2020}]{Balashev2020b}
{Balashev} S.~A.,  {Ledoux} C.,  {Noterdaeme} P.,  {Srianand} R.,  {Petitjean}
  P.,   {Gupta} N.,  2020, \mn@doi [\mnras] {10.1093/mnras/staa2108}, \href
  {https://ui.adsabs.harvard.edu/abs/2020MNRAS.497.1946B} {497, 1946}

\bibitem[\protect\citeauthoryear{{Balashev}, {Gupta}  \& {Kosenko}}{{Balashev}
  et~al.}{2021}]{Balashev2021}
{Balashev} S.~A.,  {Gupta} N.,   {Kosenko} D.~N.,  2021, \mn@doi [\mnras]
  {10.1093/mnras/stab1122}, \href
  {https://ui.adsabs.harvard.edu/abs/2021MNRAS.504.3797B} {504, 3797}

\bibitem[\protect\citeauthoryear{{Balashev}, {Telikova}  \&
  {Noterdaeme}}{{Balashev} et~al.}{2022}]{Balashev2022}
{Balashev} S.~A.,  {Telikova} K.~N.,   {Noterdaeme} P.,  2022, \mn@doi [\mnras]
  {10.1093/mnrasl/slab119}, \href
  {https://ui.adsabs.harvard.edu/abs/2022MNRAS.509L..26B} {509, L26}

\bibitem[\protect\citeauthoryear{{Barinovs}, {van Hemert}, {Krems}  \&
  {Dalgarno}}{{Barinovs} et~al.}{2005}]{Barinovs2005}
{Barinovs} {\u{G}}.,  {van Hemert} M.~C.,  {Krems} R.,   {Dalgarno} A.,  2005,
  \mn@doi [\apj] {10.1086/426860}, \href
  {https://ui.adsabs.harvard.edu/abs/2005ApJ...620..537B} {620, 537}

\bibitem[\protect\citeauthoryear{{Bell}, {Berrington}  \& {Thomas}}{{Bell}
  et~al.}{1998}]{Bell1998}
{Bell} K.~L.,  {Berrington} K.~A.,   {Thomas} M.~R.~J.,  1998, \mn@doi [\mnras]
  {10.1046/j.1365-8711.1998.01364.x}, \href
  {https://ui.adsabs.harvard.edu/abs/1998MNRAS.293L..83B} {293, L83}

\bibitem[\protect\citeauthoryear{{Bialy} \& {Sternberg}}{{Bialy} \&
  {Sternberg}}{2016}]{Bialy2016}
{Bialy} S.,  {Sternberg} A.,  2016, \mn@doi [\apj]
  {10.3847/0004-637X/822/2/83}, \href
  {https://ui.adsabs.harvard.edu/abs/2016ApJ...822...83B} {822, 83}

\bibitem[\protect\citeauthoryear{{Bialy} \& {Sternberg}}{{Bialy} \&
  {Sternberg}}{2019}]{Bialy2019}
{Bialy} S.,  {Sternberg} A.,  2019, \mn@doi [\apj] {10.3847/1538-4357/ab2fd1},
  \href {https://ui.adsabs.harvard.edu/abs/2019ApJ...881..160B} {881, 160}

\bibitem[\protect\citeauthoryear{{Fitzpatrick} \& {Massa}}{{Fitzpatrick} \&
  {Massa}}{2007}]{Fitzpatrick2007}
{Fitzpatrick} E.~L.,  {Massa} D.,  2007, \mn@doi [\apj] {10.1086/518158}, \href
  {https://ui.adsabs.harvard.edu/abs/2007ApJ...663..320F} {663, 320}

\bibitem[\protect\citeauthoryear{{Flower} \& {Launay}}{{Flower} \&
  {Launay}}{1977}]{Flower1977}
{Flower} D.~R.,  {Launay} J.~M.,  1977, \mn@doi [Journal of Physics B Atomic
  Molecular Physics] {10.1088/0022-3700/10/18/024}, \href
  {https://ui.adsabs.harvard.edu/abs/1977JPhB...10.3673F} {10, 3673}

\bibitem[\protect\citeauthoryear{Foreman-Mackey}{Foreman-Mackey}{2016}]{corner}
Foreman-Mackey D.,  2016, \mn@doi [The Journal of Open Source Software]
  {10.21105/joss.00024}, 1, 24

\bibitem[\protect\citeauthoryear{{Goodman} \& {Weare}}{{Goodman} \&
  {Weare}}{2010}]{Goodman2010}
{Goodman} J.,  {Weare} J.,  2010, \mn@doi [Communications in Applied
  Mathematics and Computational Science] {10.2140/camcos.2010.5.65}, \href
  {https://ui.adsabs.harvard.edu/abs/2010CAMCS...5...65G} {5, 65}

\bibitem[\protect\citeauthoryear{{Guimar{\~a}es}, {Noterdaeme}, {Petitjean},
  {Ledoux}, {Srianand}, {L{\'o}pez}  \& {Rahmani}}{{Guimar{\~a}es}
  et~al.}{2012}]{Guimaraes2012}
{Guimar{\~a}es} R.,  {Noterdaeme} P.,  {Petitjean} P.,  {Ledoux} C.,
  {Srianand} R.,  {L{\'o}pez} S.,   {Rahmani} H.,  2012, \mn@doi [\aj]
  {10.1088/0004-6256/143/6/147}, \href
  {https://ui.adsabs.harvard.edu/abs/2012AJ....143..147G} {143, 147}

\bibitem[\protect\citeauthoryear{{Heays}, {Bosman}  \& {van Dishoeck}}{{Heays}
  et~al.}{2017}]{Heays2017}
{Heays} A.~N.,  {Bosman} A.~D.,   {van Dishoeck} E.~F.,  2017, \mn@doi [\aap]
  {10.1051/0004-6361/201628742}, \href
  {https://ui.adsabs.harvard.edu/abs/2017A&A...602A.105H} {602, A105}

\bibitem[\protect\citeauthoryear{{Hollenbach}, {Kaufman}, {Neufeld}, {Wolfire}
  \& {Goicoechea}}{{Hollenbach} et~al.}{2012}]{Hollenbach2012}
{Hollenbach} D.,  {Kaufman} M.~J.,  {Neufeld} D.,  {Wolfire} M.,   {Goicoechea}
  J.~R.,  2012, \mn@doi [\apj] {10.1088/0004-637X/754/2/105}, \href
  {https://ui.adsabs.harvard.edu/abs/2012ApJ...754..105H} {754, 105}

\bibitem[\protect\citeauthoryear{{Indriolo} \& {McCall}}{{Indriolo} \&
  {McCall}}{2012}]{Indriolo2012}
{Indriolo} N.,  {McCall} B.~J.,  2012, \mn@doi [\apj]
  {10.1088/0004-637X/745/1/91}, \href
  {https://ui.adsabs.harvard.edu/abs/2012ApJ...745...91I} {745, 91}

\bibitem[\protect\citeauthoryear{{Indriolo} et~al.}{{Indriolo}
  et~al.}{2015}]{Indriolo2015}
{Indriolo} N.,  et~al., 2015, \mn@doi [\apj] {10.1088/0004-637X/800/1/40},
  \href {https://ui.adsabs.harvard.edu/abs/2015ApJ...800...40I} {800, 40}

\bibitem[\protect\citeauthoryear{{Ivanchik}, {Petitjean}, {Balashev},
  {Srianand}, {Varshalovich}, {Ledoux}  \& {Noterdaeme}}{{Ivanchik}
  et~al.}{2010}]{Ivanchik2010}
{Ivanchik} A.~V.,  {Petitjean} P.,  {Balashev} S.~A.,  {Srianand} R.,
  {Varshalovich} D.~A.,  {Ledoux} C.,   {Noterdaeme} P.,  2010, \mn@doi
  [\mnras] {10.1111/j.1365-2966.2010.16382.x}, \href
  {https://ui.adsabs.harvard.edu/abs/2010MNRAS.404.1583I} {404, 1583}

\bibitem[\protect\citeauthoryear{{Ivanchik}, {Balashev}, {Varshalovich}  \&
  {Klimenko}}{{Ivanchik} et~al.}{2015}]{Ivanchik2015}
{Ivanchik} A.~V.,  {Balashev} S.~A.,  {Varshalovich} D.~A.,   {Klimenko} V.~V.,
   2015, \mn@doi [Astronomy Reports] {10.1134/S1063772915020031}, \href
  {https://ui.adsabs.harvard.edu/abs/2015ARep...59..100I} {59, 100}

\bibitem[\protect\citeauthoryear{{Jaquet}, {Staemmler}, {Smith}  \&
  {Flower}}{{Jaquet} et~al.}{1992}]{Jaquet1992}
{Jaquet} R.,  {Staemmler} V.,  {Smith} M.~D.,   {Flower} D.~R.,  1992, \mn@doi
  [Journal of Physics B Atomic Molecular Physics] {10.1088/0953-4075/25/1/030},
  \href {https://ui.adsabs.harvard.edu/abs/1992JPhB...25..285J} {25, 285}

\bibitem[\protect\citeauthoryear{{Jenkins} \& {Tripp}}{{Jenkins} \&
  {Tripp}}{2001}]{Jenkins2001}
{Jenkins} E.~B.,  {Tripp} T.~M.,  2001, \mn@doi [\apjs] {10.1086/323326}, \href
  {https://ui.adsabs.harvard.edu/abs/2001ApJS..137..297J} {137, 297}

\bibitem[\protect\citeauthoryear{{Jenkins} \& {Tripp}}{{Jenkins} \&
  {Tripp}}{2011}]{Jenkins2011}
{Jenkins} E.~B.,  {Tripp} T.~M.,  2011, \mn@doi [\apj]
  {10.1088/0004-637X/734/1/65}, \href
  {https://ui.adsabs.harvard.edu/abs/2011ApJ...734...65J} {734, 65}

\bibitem[\protect\citeauthoryear{{Johnson}, {Burke}  \& {Kingston}}{{Johnson}
  et~al.}{1987}]{Johnson1987}
{Johnson} C.~T.,  {Burke} P.~G.,   {Kingston} A.~E.,  1987, \mn@doi [Journal of
  Physics B Atomic Molecular Physics] {10.1088/0022-3700/20/11/022}, \href
  {https://ui.adsabs.harvard.edu/abs/1987JPhB...20.2553J} {20, 2553}

\bibitem[\protect\citeauthoryear{{Jorgenson}, {Wolfe}  \&
  {Prochaska}}{{Jorgenson} et~al.}{2010}]{Jorgenson2010}
{Jorgenson} R.~A.,  {Wolfe} A.~M.,   {Prochaska} J.~X.,  2010, \mn@doi [\apj]
  {10.1088/0004-637X/722/1/460}, \href
  {https://ui.adsabs.harvard.edu/abs/2010ApJ...722..460J} {722, 460}

\bibitem[\protect\citeauthoryear{{Keenan}, {Lennon}, {Johnson}  \&
  {Kingston}}{{Keenan} et~al.}{1986}]{Keenan1986}
{Keenan} F.~P.,  {Lennon} D.~J.,  {Johnson} C.~T.,   {Kingston} A.~E.,  1986,
  \mn@doi [\mnras] {10.1093/mnras/220.3.571}, \href
  {https://ui.adsabs.harvard.edu/abs/1986MNRAS.220..571K} {220, 571}

\bibitem[\protect\citeauthoryear{{Klessen} \& {Glover}}{{Klessen} \&
  {Glover}}{2016}]{Klessen2016}
{Klessen} R.~S.,  {Glover} S. C.~O.,  2016, \mn@doi [Saas-Fee Advanced Course]
  {10.1007/978-3-662-47890-5\_2}, \href
  {https://ui.adsabs.harvard.edu/abs/2016SAAS...43...85K} {43, 85}

\bibitem[\protect\citeauthoryear{{Klimenko} \& {Balashev}}{{Klimenko} \&
  {Balashev}}{2020}]{Klimenko2020}
{Klimenko} V.~V.,  {Balashev} S.~A.,  2020, \mn@doi [\mnras]
  {10.1093/mnras/staa2134}, \href
  {https://ui.adsabs.harvard.edu/abs/2020MNRAS.498.1531K} {498, 1531}

\bibitem[\protect\citeauthoryear{{Klimenko}, {Balashev}, {Ivanchik}  \&
  {Varshalovich}}{{Klimenko} et~al.}{2016}]{Klimenko2016}
{Klimenko} V.~V.,  {Balashev} S.~A.,  {Ivanchik} A.~V.,   {Varshalovich} D.~A.,
   2016, \mn@doi [Astronomy Letters] {10.1134/S1063773716030038}, \href
  {https://ui.adsabs.harvard.edu/abs/2016AstL...42..137K} {42, 137}

\bibitem[\protect\citeauthoryear{{Kosenko} et~al.}{{Kosenko}
  et~al.}{2021}]{Kosenko2021}
{Kosenko} D.~N.,  et~al., 2021, \mn@doi [\mnras] {10.1093/mnras/stab1535},
  \href {https://ui.adsabs.harvard.edu/abs/2021MNRAS.505.3810K} {505, 3810}

\bibitem[\protect\citeauthoryear{{Kosenko}, {Balashev}  \&
  {Klimenko}}{{Kosenko} et~al.}{2023}]{Kosenko2023}
{Kosenko} D.~N.,  {Balashev} S.~A.,   {Klimenko} V.~V.,  2023, \mn@doi [arXiv
  e-prints] {10.48550/arXiv.2309.01599}, \href
  {https://ui.adsabs.harvard.edu/abs/2023arXiv230901599K} {p. arXiv:2309.01599}

\bibitem[\protect\citeauthoryear{{Liszt}}{{Liszt}}{2011}]{Liszt2011}
{Liszt} H.~S.,  2011, \mn@doi [\aap] {10.1051/0004-6361/201015824}, \href
  {https://ui.adsabs.harvard.edu/abs/2011A&A...527A..45L} {527, A45}

\bibitem[\protect\citeauthoryear{{Mathis}, {Mezger}  \& {Panagia}}{{Mathis}
  et~al.}{1983}]{Mathis1983}
{Mathis} J.~S.,  {Mezger} P.~G.,   {Panagia} N.,  1983, \aap, \href
  {https://ui.adsabs.harvard.edu/abs/1983A&A...128..212M} {128, 212}

\bibitem[\protect\citeauthoryear{{McElroy}, {Walsh}, {Markwick}, {Cordiner},
  {Smith}  \& {Millar}}{{McElroy} et~al.}{2013}]{McElroy2013}
{McElroy} D.,  {Walsh} C.,  {Markwick} A.~J.,  {Cordiner} M.~A.,  {Smith} K.,
  {Millar} T.~J.,  2013, \mn@doi [\aap] {10.1051/0004-6361/201220465}, \href
  {https://ui.adsabs.harvard.edu/abs/2013A&A...550A..36M} {550, A36}

\bibitem[\protect\citeauthoryear{{Milutinovic}, {Ellison}, {Prochaska}  \&
  {Tumlinson}}{{Milutinovic} et~al.}{2010}]{Milutinovic2010}
{Milutinovic} N.,  {Ellison} S.~L.,  {Prochaska} J.~X.,   {Tumlinson} J.,
  2010, \mn@doi [\mnras] {10.1111/j.1365-2966.2010.17280.x}, \href
  {https://ui.adsabs.harvard.edu/abs/2010MNRAS.408.2071M} {408, 2071}

\bibitem[\protect\citeauthoryear{{Monteiro} \& {Flower}}{{Monteiro} \&
  {Flower}}{1987}]{Monteiro1987}
{Monteiro} T.~S.,  {Flower} D.~R.,  1987, \mn@doi [\mnras]
  {10.1093/mnras/228.1.101}, \href
  {https://ui.adsabs.harvard.edu/abs/1987MNRAS.228..101M} {228, 101}

\bibitem[\protect\citeauthoryear{{Noterdaeme}, {Ledoux}, {Petitjean}, {Le
  Petit}, {Srianand}  \& {Smette}}{{Noterdaeme} et~al.}{2007}]{Noterdaeme2007}
{Noterdaeme} P.,  {Ledoux} C.,  {Petitjean} P.,  {Le Petit} F.,  {Srianand} R.,
    {Smette} A.,  2007, \mn@doi [\aap] {10.1051/0004-6361:20078021}, \href
  {https://ui.adsabs.harvard.edu/abs/2007A&A...474..393N} {474, 393}

\bibitem[\protect\citeauthoryear{{Noterdaeme}, {Ledoux}, {Petitjean}  \&
  {Srianand}}{{Noterdaeme} et~al.}{2008a}]{Noterdaeme2008a}
{Noterdaeme} P.,  {Ledoux} C.,  {Petitjean} P.,   {Srianand} R.,  2008a,
  \mn@doi [\aap] {10.1051/0004-6361:20078780}, \href
  {https://ui.adsabs.harvard.edu/abs/2008A&A...481..327N} {481, 327}

\bibitem[\protect\citeauthoryear{{Noterdaeme}, {Petitjean}, {Ledoux},
  {Srianand}  \& {Ivanchik}}{{Noterdaeme} et~al.}{2008b}]{Noterdaeme2008b}
{Noterdaeme} P.,  {Petitjean} P.,  {Ledoux} C.,  {Srianand} R.,   {Ivanchik}
  A.,  2008b, \mn@doi [\aap] {10.1051/0004-6361:200810414}, \href
  {https://ui.adsabs.harvard.edu/abs/2008A&A...491..397N} {491, 397}

\bibitem[\protect\citeauthoryear{{Noterdaeme}, {Petitjean}, {Ledoux},
  {L{\'o}pez}, {Srianand}  \& {Vergani}}{{Noterdaeme}
  et~al.}{2010}]{Noterdaeme2010}
{Noterdaeme} P.,  {Petitjean} P.,  {Ledoux} C.,  {L{\'o}pez} S.,  {Srianand}
  R.,   {Vergani} S.~D.,  2010, \mn@doi [\aap] {10.1051/0004-6361/201015147},
  \href {https://ui.adsabs.harvard.edu/abs/2010A&A...523A..80N} {523, A80}

\bibitem[\protect\citeauthoryear{{Noterdaeme}, {Srianand}, {Rahmani},
  {Petitjean}, {P{\^a}ris}, {Ledoux}, {Gupta}  \& {L{\'o}pez}}{{Noterdaeme}
  et~al.}{2015}]{Noterdaeme2015}
{Noterdaeme} P.,  {Srianand} R.,  {Rahmani} H.,  {Petitjean} P.,  {P{\^a}ris}
  I.,  {Ledoux} C.,  {Gupta} N.,   {L{\'o}pez} S.,  2015, \mn@doi [\aap]
  {10.1051/0004-6361/201425376}, \href
  {https://ui.adsabs.harvard.edu/abs/2015A&A...577A..24N} {577, A24}

\bibitem[\protect\citeauthoryear{{Noterdaeme} et~al.,}{{Noterdaeme}
  et~al.}{2017}]{Noterdaeme2017}
{Noterdaeme} P.,  et~al., 2017, \mn@doi [\aap] {10.1051/0004-6361/201629173},
  \href {https://ui.adsabs.harvard.edu/abs/2017A&A...597A..82N} {597, A82}

\bibitem[\protect\citeauthoryear{{Noterdaeme}, {Ledoux}, {Zou}, {Petitjean},
  {Srianand}, {Balashev}  \& {L{\'o}pez}}{{Noterdaeme}
  et~al.}{2018}]{Noterdaeme2018}
{Noterdaeme} P.,  {Ledoux} C.,  {Zou} S.,  {Petitjean} P.,  {Srianand} R.,
  {Balashev} S.,   {L{\'o}pez} S.,  2018, \mn@doi [\aap]
  {10.1051/0004-6361/201732266}, \href
  {https://ui.adsabs.harvard.edu/abs/2018A&A...612A..58N} {612, A58}

\bibitem[\protect\citeauthoryear{{Noterdaeme}, {Balashev}, {Krogager},
  {Srianand}, {Fathivavsari}, {Petitjean}  \& {Ledoux}}{{Noterdaeme}
  et~al.}{2019}]{Noterdaeme2019}
{Noterdaeme} P.,  {Balashev} S.,  {Krogager} J.~K.,  {Srianand} R.,
  {Fathivavsari} H.,  {Petitjean} P.,   {Ledoux} C.,  2019, \mn@doi [\aap]
  {10.1051/0004-6361/201935371}, \href
  {https://ui.adsabs.harvard.edu/abs/2019A&A...627A..32N} {627, A32}

\bibitem[\protect\citeauthoryear{{Noterdaeme}, {Balashev}, {Krogager},
  {Laursen}, {Srianand}, {Gupta}, {Petitjean}  \& {Fynbo}}{{Noterdaeme}
  et~al.}{2021}]{Noterdaeme2021}
{Noterdaeme} P.,  {Balashev} S.,  {Krogager} J.~K.,  {Laursen} P.,  {Srianand}
  R.,  {Gupta} N.,  {Petitjean} P.,   {Fynbo} J.~P.~U.,  2021, \mn@doi [\aap]
  {10.1051/0004-6361/202038877}, \href
  {https://ui.adsabs.harvard.edu/abs/2021A&A...646A.108N} {646, A108}

\bibitem[\protect\citeauthoryear{{Noterdaeme} et~al.,}{{Noterdaeme}
  et~al.}{2023}]{Noterdaeme2023}
{Noterdaeme} P.,  et~al., 2023, \mn@doi [\aap] {10.1051/0004-6361/202245554},
  \href {https://ui.adsabs.harvard.edu/abs/2023A&A...673A..89N} {673, A89}

\bibitem[\protect\citeauthoryear{{Ranjan} et~al.,}{{Ranjan}
  et~al.}{2018}]{Ranjan2018}
{Ranjan} A.,  et~al., 2018, \mn@doi [\aap] {10.1051/0004-6361/201833446}, \href
  {https://ui.adsabs.harvard.edu/abs/2018A&A...618A.184R} {618, A184}

\bibitem[\protect\citeauthoryear{{R{\'e}my-Ruyer} et~al.}{{R{\'e}my-Ruyer}
  et~al.}{2014}]{RemyRuyer2014}
{R{\'e}my-Ruyer} A.,  et~al., 2014, \mn@doi [\aap]
  {10.1051/0004-6361/201322803}, \href
  {https://ui.adsabs.harvard.edu/abs/2014A&A...563A..31R} {563, A31}

\bibitem[\protect\citeauthoryear{{Roman-Duval} et~al.,}{{Roman-Duval}
  et~al.}{2019}]{RomanDuval2019}
{Roman-Duval} J.,  et~al., 2019, \mn@doi [\apj] {10.3847/1538-4357/aaf8bb},
  \href {https://ui.adsabs.harvard.edu/abs/2019ApJ...871..151R} {871, 151}

\bibitem[\protect\citeauthoryear{{Schroder}, {Staemmler}, {Smith}, {Flower}  \&
  {Jaquet}}{{Schroder} et~al.}{1991}]{Schroder1991}
{Schroder} K.,  {Staemmler} V.,  {Smith} M.~D.,  {Flower} D.~R.,   {Jaquet} R.,
   1991, \mn@doi [Journal of Physics B Atomic Molecular Physics]
  {10.1088/0953-4075/24/10/007}, \href
  {https://ui.adsabs.harvard.edu/abs/1991JPhB...24.2487S} {24, 2487}

\bibitem[\protect\citeauthoryear{{Silva} \& {Viegas}}{{Silva} \&
  {Viegas}}{2002}]{Silva2002}
{Silva} A.~I.,  {Viegas} S.~M.,  2002, \mn@doi [\mnras]
  {10.1046/j.1365-8711.2002.04956.x}, \href
  {https://ui.adsabs.harvard.edu/abs/2002MNRAS.329..135S} {329, 135}

\bibitem[\protect\citeauthoryear{{Spitzer}}{{Spitzer}}{1978}]{Spitzer1978}
{Spitzer} L.,  1978, {Physical processes in the interstellar medium},
  \mn@doi{10.1002/9783527617722.
}

\bibitem[\protect\citeauthoryear{{Srianand}, {Petitjean}, {Ledoux}, {Ferland}
  \& {Shaw}}{{Srianand} et~al.}{2005}]{Srianand2005}
{Srianand} R.,  {Petitjean} P.,  {Ledoux} C.,  {Ferland} G.,   {Shaw} G.,
  2005, \mn@doi [\mnras] {10.1111/j.1365-2966.2005.09324.x}, \href
  {https://ui.adsabs.harvard.edu/abs/2005MNRAS.362..549S} {362, 549}

\bibitem[\protect\citeauthoryear{{Srianand}, {Noterdaeme}, {Ledoux}  \&
  {Petitjean}}{{Srianand} et~al.}{2008}]{Srianand2008}
{Srianand} R.,  {Noterdaeme} P.,  {Ledoux} C.,   {Petitjean} P.,  2008, \mn@doi
  [\aap] {10.1051/0004-6361:200809727}, \href
  {https://ui.adsabs.harvard.edu/abs/2008A&A...482L..39S} {482, L39}

\bibitem[\protect\citeauthoryear{{Staemmler} \& {Flower}}{{Staemmler} \&
  {Flower}}{1991}]{Staemmler1991}
{Staemmler} V.,  {Flower} D.~R.,  1991, \mn@doi [Journal of Physics B Atomic
  Molecular Physics] {10.1088/0953-4075/24/9/013}, \href
  {https://ui.adsabs.harvard.edu/abs/1991JPhB...24.2343S} {24, 2343}

\bibitem[\protect\citeauthoryear{{Sternberg}, {Le Petit}, {Roueff}  \& {Le
  Bourlot}}{{Sternberg} et~al.}{2014}]{Sternberg2014}
{Sternberg} A.,  {Le Petit} F.,  {Roueff} E.,   {Le Bourlot} J.,  2014, \mn@doi
  [\apj] {10.1088/0004-637X/790/1/10}, \href
  {https://ui.adsabs.harvard.edu/abs/2014ApJ...790...10S} {790, 10}

\bibitem[\protect\citeauthoryear{{Wolfire}, {McKee}, {Hollenbach}  \&
  {Tielens}}{{Wolfire} et~al.}{2003}]{Wolfire2003}
{Wolfire} M.~G.,  {McKee} C.~F.,  {Hollenbach} D.,   {Tielens} A.~G.~G.~M.,
  2003, \mn@doi [\apj] {10.1086/368016}, \href
  {https://ui.adsabs.harvard.edu/abs/2003ApJ...587..278W} {587, 278}

\bibitem[\protect\citeauthoryear{{Wolfire}, {Tielens}, {Hollenbach}  \&
  {Kaufman}}{{Wolfire} et~al.}{2008}]{Wolfire2008}
{Wolfire} M.~G.,  {Tielens} A.~G.~G.~M.,  {Hollenbach} D.,   {Kaufman} M.~J.,
  2008, \mn@doi [\apj] {10.1086/587688}, \href
  {https://ui.adsabs.harvard.edu/abs/2008ApJ...680..384W} {680, 384}

\makeatother
\end{thebibliography}




\appendix
\bsp	
\label{lastpage}
\end{document}